\documentclass[twocolumn, times, tighten]{aastex63}

\newcommand{\elv}{$E(\lambda-V)$}
\newcommand{\ebv}{$E(B-V)$}
\newcommand{\alav}{$A(\lambda)/A(V)$}
\newcommand{\elkejk}{$E(\lambda-K)/E(J-K)$}
\newcommand{\alak}{$A(\lambda)/A(K)$}

\usepackage{footmisc}

\begin{document}

\shortauthors{Gordon et al.}
\shorttitle{MIR Extinction}

\title{Milky Way Mid-Infrared {\it Spitzer} Spectroscopic Extinction Curves: Continuum and Silicate Features}

\author[0000-0001-5340-6774]{Karl D.\ Gordon}
\affiliation{Space Telescope Science Institute, 3700 San Martin
  Drive, Baltimore, MD, 21218, USA}
\affiliation{Sterrenkundig Observatorium, Universiteit Gent,
  Gent, Belgium}

\author{Karl A.\ Misselt}
\affiliation{Steward Observatory, University of Arizona, Tucson,
  AZ 85721, USA}

\author[0000-0003-4757-2500]{Jeroen Bouwman}
\affiliation{Max-Planck-Institut f\"ur Astronomie, K\"onigstuhl 17,
  D-69117 Heidelberg, Germany}

\author[0000-0002-0141-7436]{Geoffrey~C.~Clayton}
\affiliation{Department of Physics \& Astronomy, Louisiana State University,
  Baton Rouge, LA 70803, USA}

\author[0000-0001-9462-5543]{Marjorie Decleir}
\affiliation{Space Telescope Science Institute, 3700 San Martin
  Drive, Baltimore, MD, 21218, USA}

\author{Dean~C.\ Hines}
\affiliation{Space Telescope Science Institute, 3700 San Martin
  Drive, Baltimore, MD, 21218, USA}

\author[0000-0001-8102-2903]{Yvonne~Pendleton}
\affiliation{NASA Ames Research Center, Moffett Field, CA 94035, USA}

\author{George~Rieke}
\affiliation{Steward Observatory, University of Arizona, Tucson,
  AZ 85721, USA}

\author[0000-0003-1545-5078]{J.~D.~T.~Smith}
\affiliation{Ritter Astrophysical Research Center, University of Toledo,
  Toledo, OH 43606, USA}

\author[0000-0001-8539-3891]{D.~C.~B.~Whittet}
\affiliation{Department of Physics, Applied Physics and Astronomy,
  Rensselaer Polytechnic Institute, 110 Eighth Street, Troy, NY 12180, USA}

\begin{abstract}
We measured the mid-infrared (MIR) extinction using {\it Spitzer} photometry and spectroscopy (3.6--37~\micron) for a sample of Milky Way sightlines (mostly) having measured ultraviolet extinction curves.
We used the pair method to determine the MIR extinction that we then fit with a power law for the continuum and modified Drude profiles for the silicate features.
We derived 16 extinction curves having a range of $A(V)$ (1.8--5.5) and $R(V)$ values (2.4--4.3).
Our sample includes two dense sightlines that have 3~\micron\ ice feature detections and weak 2175~\AA\ bumps.
The average $A(\lambda)/A(V)$ diffuse sightline extinction curve we calculate is lower than most previous literature measurements.
This agrees better with literature diffuse dust grain models, though it is somewhat higher.
The 10~\micron\ silicate feature does not correlate with the 2175~\AA\ bump, for the first time providing direct observational confirmation that these two features arise from different grain populations.
The strength of the 10~\micron\ silicate feature varies by $\sim$2.5 and is not correlated with $A(V)$ or $R(V)$.
It is well fit by a modified Drude profile with strong correlations seen between the central wavelength, width, and asymmetry.
We do not detect other features with limits in $A(\lambda)/A(V)$ units of 0.0026 (5--10~\micron), 0.004 (10--20~\micron), and 0.008 (20-40~\micron).
We find that the standard prescription of estimating $R(V)$ from $C \times E(K_s-V)/E(B-V)$ has $C = -1.14$ and a scatter of $\sim$7\%.
Using the IRAC 5.6~\micron\ band instead of $K_s$ gives $C = -1.03$ and the least scatter of $\sim$3\%.
\end{abstract}


\section{Introduction}
\label{sec_intro}

Extinction curves are one of the fundamental tools for understanding both the nature of interstellar medium (ISM) dust and its effects on observations of obscured sources.
Models of interstellar dust \citep[e.g.,][]{Weingartner01, Clayton03, Zubko04, Jones13} use them as one of the basic constraints on the size and composition of dust grains.
Unlike many other methods of investigating the nature of dust in the ISM, extinction curves offer a direct probe of the physical properties of dust, unaffected by, for example, geometry or radiation fields.
It is well understood that there are significant variations in the ultraviolet/optical \citep{Cardelli89, Valencic04, Gordon09FUSE, Fitzpatrick19} and mid-infrared \citep{Rieke89, Indebetouw05, Chiar06, Zasowski09, Fritz11, Xue16} extinction curves, differences largely attributed to changes in grain properties with environment.
Understanding these variations and their connections across a wide wavelength range and in different environments is critical to understanding how ISM dust absorbs and emits radiation.

The measurement of an extinction curve is, in principle, straightforward.
Given two identical sources, one having no dust along the line of sight and one located behind a column of dust, the wavelength dependence of the ratio of the fluxes from the sources measures only the dust signature, as the source spectrum is canceled out.
For extinction curve work, a straightforward way to find two identical sources is to use two stars with the same spectral type and metallicity, one showing no or a small amount of reddening and one showing significant reddening.
Identical spectral types and metallicities indicate the same surface physics for each star (e.g., same effective temperature and surface gravity) and, therefore, the stars should have identical spectra.
The relative paucity of lines in the spectra of O and B main sequence stars coupled with their overall brightness makes them ideal candidates for use in measuring the extinction curve from the ultraviolet to the mid-infrared.

The general shape of Milky Way extinction curves has been well established from the ultraviolet (UV) to near-infrared (NIR, $\sim$1-5~\micron) \citep{Cardelli89, Martin90, Valencic04, Fitzpatrick09, Gordon09FUSE, Fitzpatrick19}.
The wavelength dependence of the extinction curve probes the size distribution (through the shape of the curve; the interaction of photons with grains is most effective when $\lambda \sim a$, where $a$ is the diameter of the grain) as well as the composition of ISM grains (through the presence of features, e.g., the 2175~\AA\ bump).
Not surprisingly, there are significant variations in the extinction curve, both in overall shape and the strength of features, along lines of sight through varied environments.
A large fraction of these variations can be represented by a single parameter, chosen to be $R(V) = A(V)/E(B-V)$ \citep{Cardelli89, Valencic04, Gordon09FUSE, Fitzpatrick19}.
As the grains that dominate the extinction in the UV versus the optical and NIR are of different sizes, this result suggests that the variations in the abundances of grains of different sizes are not independent but are related through the environmental process modifying the overall size distribution.

Measuring the extinction beyond 5~\micron\ and its dependence on environment is important as such extinction probes large dust grains in general and silicate dust grains of all sizes.
The silicate grains are directly probed as they have strong absorption features around 10 and 20~\micron.
The contribution of large dust grains is uniquely measured by the overall shape of the mid-infrared (MIR) extinction curve as such grains produce little differential extinction at UV or optical wavelengths.
Several lines of evidence point to the possible abundance of large grains in the interstellar medium.
The most direct argument comes from in situ measurements of interstellar dust grains entering the solar system from the local ISM, where it is found that the majority of the mass resides in grains larger than 1~$\micron$ \citep{Frisch99}.
Furthermore, a significant population of large dust grains with very low optical extinction is indicated in analyses of dense cloud dust extinction efficiencies \citep{Kim96}.
The analysis of MIR images of dense clouds has revealed a strong scattered light signal that has been identified as due to scattering from large micron sized dust grains \citep{Steinacker10, Steinacker15, Ysard16}.

Observations have been made quantifying extinction in the 5-13~$\micron$ region based on ground-based data that includes the 10~\micron\ silicate feature \citep{Roche84, Roche85, Rieke85, Rieke89, Bowey98, Bowey04} and even to longer wavelengths using the Infrared Space Observatory observations that also include the 20~$\micron$ silicate feature \citep{Schutte98, Lutz99, Kemper04, Chiar06, Jiang06, Fritz11}.
In these cases, the majority of the sightlines studied have used extreme stars (e.g., hypergiants, Wolf-Rayet stars, etc.) often as they were very bright and allowed measurements of high dust columns,  yet are unsuitable for precise extinction curve work due to difficulties in determining their intrinsic spectra.
These works have found variations in the MIR, possibly indicating that the diffuse and dense ISM regions have different dust properties.

A number of extinction studies using Spitzer or WISE photometric data have been carried out, focusing on using photometry of large samples of stars to measure the average MIR extinction curve.
These studies can be broken into those that focus on the extinction towards the Galactic Center \citep{Nishiyama09}, towards molecular clouds and star-forming regions \citep{Flaherty07, Chapman09, Ascenso13, Wang13}, and towards the Galactic plane as probed by the GLIMPSE/MIPSGAL and other surveys \citep{Indebetouw05, Gao09, Zasowski09, Xue16, Wang17}.
In general, these studies are most sensitive to the extinction in regions of high optical depth.
Some studies find no evidence for variations in the MIR extinction \citep{Ascenso13}, while others do see variations \citep{Chapman09, Gao09, Zasowski09, Wang13, Wang17}.

Using Spitzer IRS spectroscopy, a few sightlines have been investigated mainly focusing on the 10~\micron\ silicate feature.
For five sightlines, \citet{Shao18} found that the 10~\micron\ silicate feature central wavelength and width vary and these variations are not correlated with specific environment.
\citet{Hensley20} carried out a detailed study of the sightline towards Cyg~OB2~12 (aka VI~Cyg~12) deriving an extinction curve that utilizes other studies for the absolute normalization.  In addition, they find three small absorption features in the 6--8~\micron\ region they identify with aromatic and aliphatic materials although \citet{Potapov20} do not find as many features in their analysis of the same data.
Cyg~OB2~12 is often used to study diffuse dust extinction in the IR as it has no 3~\micron\ ice feature even though it has an $A(V) \sim 10$ \citep{Whittet15}.

We undertook a program to measure MIR extinction curves in the Milky Way at spectroscopic resolution for sightlines with existing UV extinction curves.
One of the prime goals is to directly connect measurements of dust properties probed in the UV (2175~\AA\ bump, small grain extinction) and MIR (silicate features).
Another goal is to derive MIR extinction curves for smaller dust columns than are typically measured in the MIR, better ensuring that truly diffuse dust would be studied.
Two Spitzer programs were carried out to obtain spectroscopic (IRS spectra) and photometric (IRAC, IRS blue peakup, and MIPS 24~\micron) data of a representative sample of sightlines towards OB stars.
Section~\ref{sec_data} gives the sample and describes the data used in this study with visualizations of the IRS apertures superimposed on the MIPS24 images given in Appendix~\ref{spitzer_images}.
The calculation, normalization, and fitting of the MIR dust extinction curves are presented in \S\ref{sec_curves}.
Section~\ref{sec_results} gives the average diffuse extinction, how it compares to previous measurements and dust grain models, the variation in the 10 \& 20~\micron\ silicate feature properties in our sample, limits on the strength of other extinction features, and prescriptions for deriving $R(V)$ and $A(V)$.
Finally, we give the conclusions of our work in \S\ref{sec_conclusions}.

\section{Data}
\label{sec_data}

\begin{deluxetable*}{lcccccccc}[tbp]
\tablewidth{0pt}
\tabletypesize{\footnotesize}
\tablecaption{Spitzer Data\label{tab_phot}}
\tablehead{\colhead{Name} & \colhead{IRAC1} & \colhead{IRAC2} & \colhead{IRAC3} & \colhead{IRAC4} & \colhead{IRSB} & \colhead{MIPS24} & \colhead{IRS} & \colhead{Extraction} \\
& \colhead{[mag]} & \colhead{[mag]} & \colhead{[mag]} & \colhead{[mag]} & \colhead{[mag]} & \colhead{[mag]} & \colhead{[$\micron$]}}
\startdata
\multicolumn{8}{c}{Comparison Stars} \\ \hline
HD031726 & $ 6.798 \pm  0.044$ & $ 6.836 \pm  0.044$ & $ 6.752 \pm  0.044$ & $ 6.846 \pm  0.044$ & $ 6.887 \pm  0.045$ & $ 6.957 \pm  0.055$ & 5-33 & SMART \\
HD034816 & $ 5.040 \pm  0.044$ & $ 5.095 \pm  0.044$ & $ 5.000 \pm  0.046$ & $ 5.104 \pm  0.044$ & $ 5.172 \pm  0.043$ & $ 5.181 \pm  0.047$ & 5-33 & SMART \\
HD036512 & $ 5.429 \pm  0.044$ & $ 5.470 \pm  0.044$ & $ 5.396 \pm  0.049$ & $ 5.473 \pm  0.045$ & $ 5.563 \pm  0.045$ & $ 5.603 \pm  0.050$ & 5-32 & SMART \\
HD047839 & $ 5.376 \pm  0.043$ & $ 5.423 \pm  0.044$ & $ 5.313 \pm  0.045$ & $ 5.415 \pm  0.044$ & $ 5.433 \pm  0.113$ & $ 5.670 \pm  0.053$ & 5-21 & SMART \\
HD051283 & $ 5.618 \pm  0.044$ & $ 5.636 \pm  0.044$ & $ 5.443 \pm  0.050$ & $ 5.487 \pm  0.046$ & $ 5.277 \pm  0.044$ & $ 5.154 \pm  0.047$ & 5-37 & SMART \\
HD064760 & $ 4.530 \pm  0.043$ & $ 4.558 \pm  0.044$ & $ 4.443 \pm  0.044$ & $ 4.494 \pm  0.044$ & $ 4.475 \pm  0.043$ & $ 4.419 \pm  0.048$  & 5-30 & SMART \\
HD064802 & $ 5.979 \pm  0.044$ & $ 6.005 \pm  0.044$ & $ 5.897 \pm  0.052$ & $ 5.971 \pm  0.047$ & $ 5.771 \pm  0.044$ & $ 5.971 \pm  0.055$  & 5-34 & SMART \\
HD074273 & $ 6.433 \pm  0.044$ & $ 6.496 \pm  0.046$ & $ 6.369 \pm  0.065$ & $ 6.513 \pm  0.053$ & $ 6.463 \pm  0.045$ & $ 6.605 \pm  0.053$  & 5-37 & SMART \\
HD165024 & $ 3.881 \pm  0.044$ & $ 3.932 \pm  0.043$ & $ 3.817 \pm  0.044$ & $ 3.889 \pm  0.044$ & $ 3.884 \pm  0.043$ & $ 3.861 \pm  0.045$  & 5-33 & SMART \\
HD188209 & $ 5.831 \pm  0.044$ & $ 5.854 \pm  0.044$ & $ 5.717 \pm  0.054$ & $ 5.787 \pm  0.047$ & $ 5.702 \pm  0.043$ & $ 5.426 \pm  0.047$  & 5-34 & SMART \\
HD195986 & $ 6.823 \pm  0.044$ & $ 6.856 \pm  0.045$ & $ 6.747 \pm  0.058$ & $ 6.697 \pm  0.048$ & $ 5.651 \pm  0.288$ & $ 6.892 \pm  0.056$  & 5-21 & SMART \\
HD204172 & $ 6.140 \pm  0.044$ & $ 6.161 \pm  0.045$ & $ 6.075 \pm  0.056$ & $ 6.108 \pm  0.051$ & $ 6.026 \pm  0.044$ & $ 5.934 \pm  0.050$  & 5-37 & SMART \\
HD214680 & $ 5.516 \pm  0.044$ & $ 5.559 \pm  0.044$ & $ 5.470 \pm  0.049$ & $ 5.604 \pm  0.045$ & $ 5.668 \pm  0.044$ & $ 5.748 \pm  0.050$  & 5-34 & SMART \\ \hline
\multicolumn{8}{c}{Reddened Stars} \\ \hline
BD+63D1964 & $ 6.350 \pm  0.044$ & $ 6.303 \pm  0.044$ & $ 6.127 \pm  0.047$ & $ 6.170 \pm  0.045$ & $ 5.922 \pm  0.044$ & $ 5.829 \pm  0.052$  & 5-22 & custom \\
HD014956 & $ 5.159 \pm  0.044$ & $ 5.108 \pm  0.044$ & $ 4.942 \pm  0.046$ & $ 4.965 \pm  0.044$ & $ 4.962 \pm  0.043$ & $ 4.813 \pm  0.050$  & 5-32 & SMART \\
HD029309 & $ 5.839 \pm  0.044$ & $ 5.846 \pm  0.044$ & $ 5.704 \pm  0.051$ & $ 5.757 \pm  0.047$ & $ 5.715 \pm  0.044$ & $ 5.736 \pm  0.052$  & 5-33 & SMART \\
HD029647 & $ 5.202 \pm  0.043$ & $ 5.172 \pm  0.043$ & $ 5.030 \pm  0.044$ & $ 5.104 \pm  0.044$ & $ 4.223 \pm  0.045$ & $ 5.033 \pm  0.061$  & 5-21 & SMART \\
HD034921 & $ 5.739 \pm  0.043$ & $ 5.580 \pm  0.044$ & $ 5.264 \pm  0.048$ & $ 5.063 \pm  0.045$ & $ 4.230 \pm  0.044$ & $ 4.084 \pm  0.047$  & 5-37 & SMART \\
HD096042 & $ 8.076 \pm  0.044$ & $ 8.101 \pm  0.044$ & $ 7.978 \pm  0.049$ & $ 7.992 \pm  0.046$ & $ 4.391 \pm  0.103$ & \nodata  & 5-21 & SMART \\
HD112272 & $ 4.932 \pm  0.043$ & $ 4.915 \pm  0.044$ & $ 4.720 \pm  0.045$ & $ 4.782 \pm  0.044$ & $ 4.547 \pm  0.044$ & $ 4.587 \pm  0.046$  & 5-34 & custom \\
HD147701 & $ 6.030 \pm  0.044$ & $ 6.021 \pm  0.045$ & $ 5.944 \pm  0.055$ & $ 6.011 \pm  0.048$ & $ 6.059 \pm  0.047$ & $ 6.113 \pm  0.051$  & 5-34 & custom \\
HD147889 & $ 4.319 \pm  0.044$ & $ 4.293 \pm  0.043$ & $ 4.138 \pm  0.044$ & $ 4.197 \pm  0.044$ & \nodata & $ 4.045 \pm  0.051$  & 5-19 & SMART \\
HD147933 & $ 3.469 \pm  0.043$ & $ 3.518 \pm  0.044$ & $ 3.420 \pm  0.044$ & $ 3.535 \pm  0.044$ & $ 3.488 \pm  0.044$ & $ 3.546 \pm  0.045$  & 5-32 & custom \\
HD149038 & $ 4.523 \pm  0.043$ & $ 4.538 \pm  0.044$ & $ 4.380 \pm  0.045$ & $ 4.428 \pm  0.044$ & $ 4.206 \pm  0.044$ & $ 4.168 \pm  0.046$  & \nodata & \nodata \\
HD149404 & $ 4.008 \pm  0.043$ & $ 3.918 \pm  0.044$ & $ 3.774 \pm  0.044$ & $ 3.759 \pm  0.044$ & $ 3.471 \pm  0.044$ & $ 3.194 \pm  0.048$  & 5-37 & SMART \\
HD152408 & $ 4.661 \pm  0.043$ & $ 4.522 \pm  0.043$ & $ 4.283 \pm  0.044$ & $ 4.121 \pm  0.044$ & $ 3.506 \pm  0.044$ & $ 3.117 \pm  0.045$  & 5-37 & SMART \\
HD166734 & $ 5.074 \pm  0.044$ & $ 5.011 \pm  0.044$ & $ 4.825 \pm  0.046$ & $ 4.782 \pm  0.044$ & $ 4.449 \pm  0.044$ & $ 4.105 \pm  0.048$  & 5-37 & SMART \\
HD169454 & $ 3.629 \pm  0.043$ & $ 3.537 \pm  0.043$ & $ 3.348 \pm  0.044$ & $ 3.329 \pm  0.044$ & \nodata & $ 2.773 \pm  0.048$  & 5-21 & SMART \\
HD192660 & $ 5.723 \pm  0.044$ & $ 5.711 \pm  0.044$ & $ 5.548 \pm  0.050$ & $ 5.598 \pm  0.046$ & $ 5.455 \pm  0.044$ & $ 5.412 \pm  0.049$  & 5-32 & SMART \\
HD197702 & $ 7.443 \pm  0.044$ & $ 7.456 \pm  0.044$ & $ 7.292 \pm  0.045$ & $ 7.356 \pm  0.044$ & $ 6.313 \pm  0.045$ & $ 6.262 \pm  0.052$  & 5-34 & SMART \\
HD204827 & $ 6.217 \pm  0.044$ & $ 6.226 \pm  0.046$ & $ 6.121 \pm  0.059$ & $ 6.277 \pm  0.051$ & $ 6.278 \pm  0.044$ & $ 6.308 \pm  0.052$  & 5-32 & custom \\
HD206773 & $ 5.921 \pm  0.044$ & $ 5.814 \pm  0.044$ & $ 5.570 \pm  0.050$ & $ 5.456 \pm  0.045$ & $ 4.692 \pm  0.045$ & $ 4.384 \pm  0.047$  & 5-37 & SMART \\
HD229059 & $ 4.440 \pm  0.043$ & $ 4.322 \pm  0.043$ & $ 4.107 \pm  0.044$ & $ 4.058 \pm  0.043$ & \nodata & $ 3.325 \pm  0.048$  & 5-31 & SMART \\
HD229238 & $ 6.595 \pm  0.044$ & $ 6.599 \pm  0.043$ & $ 6.423 \pm  0.044$ & $ 6.484 \pm  0.044$ & $ 6.197 \pm  0.085$ & $ 6.329 \pm  0.054$  & 5-32 & SMART \\
HD281159 & $ 6.379 \pm  0.043$ & $ 6.375 \pm  0.044$ & $ 6.209 \pm  0.049$ & $ 6.128 \pm  0.045$ & $ 2.966 \pm  0.135$ & $ 2.892 \pm  0.044$  & 5-14 & SMART \\
HD283809 & $ 5.912 \pm  0.044$ & $ 5.875 \pm  0.045$ & $ 5.727 \pm  0.051$ & $ 5.849 \pm  0.047$ & $ 5.812 \pm  0.045$ & $ 5.971 \pm  0.061$  & 5-35 & SMART \\
CYGOB2-1 & $ 7.152 \pm  0.044$ & $ 7.139 \pm  0.045$ & $ 7.056 \pm  0.064$ & $ 6.989 \pm  0.050$ & $ 7.759 \pm  0.121$ & $ 7.382 \pm  0.114$  & 5-21 & SMART \\
CYGOB2-2 & $ 7.398 \pm  0.044$ & $ 7.398 \pm  0.046$ & $ 7.343 \pm  0.078$ & $ 7.231 \pm  0.054$ & $ 7.882 \pm  0.113$ & $ 7.831 \pm  0.115$  & 5-21 & SMART \\
CYGOB2-5 & $ 4.146 \pm  0.043$ & $ 3.972 \pm  0.043$ & $ 3.752 \pm  0.043$ & $ 3.618 \pm  0.043$ & \nodata & $ 2.753 \pm  0.046$  & 5-21 & SMART \\
CYG8OB2-A & $ 5.257 \pm  0.044$ & $ 5.209 \pm  0.043$ & $ 5.041 \pm  0.044$ & $ 5.046 \pm  0.044$ & $ 4.462 \pm  0.397$ & \nodata  & 5-16 & SMART
\enddata
\end{deluxetable*}

\begin{deluxetable*}{lcccccccc}[tbp]
\tablewidth{0pt}
\tablecaption{Optical/NIR Photometry\label{tab_ophot}}
\tablehead{\colhead{Name} & \colhead{U} & \colhead{B} & \colhead{V} & \colhead{Ref} &
    \colhead{J} & \colhead{H} & \colhead{K$_\mathrm{s}$} & \colhead{Ref}}
\startdata
\multicolumn{9}{c}{Comparison Stars} \\ \hline
HD031726 & $ 5.068 \pm  0.005$ & $ 5.941 \pm  0.002$ & $ 6.147 \pm  0.001$ & 1 &
    $ 6.642 \pm  0.020$ & $ 6.779 \pm  0.047$ & $ 6.803 \pm  0.024$ & A \\
HD034816 & $ 3.030 \pm  0.035$ & $ 4.040 \pm  0.028$ & $ 4.290 \pm  0.020$ & 2 &
    $ 4.882 \pm  0.018$ & $ 5.001 \pm  0.047$ & $ 5.004 \pm  0.017$ & A \\
HD036512 & $ 3.290 \pm  0.035$ & $ 4.360 \pm  0.028$ & $ 4.620 \pm  0.020$ & 2 &
    $ 5.363 \pm  0.037$ & $ 5.347 \pm  0.038$ & $ 5.376 \pm  0.023$ & A \\
HD047839 & $ 3.350 \pm  0.020$ & $ 4.420 \pm  0.020$ & $ 4.660 \pm  0.020$ & 2 &
    $ 5.202 \pm  0.023$ & $ 5.322 \pm  0.021$ & $ 5.340 \pm  0.021$ & A \\
HD051283 & $ 4.190 \pm  0.040$ & $ 5.150 \pm  0.029$ & $ 5.320 \pm  0.023$ & 1 &
    $ 5.632 \pm  0.023$ & $ 5.653 \pm  0.036$ & $ 5.680 \pm  0.020$ & A \\
HD064760 & $ 3.110 \pm  0.035$ & $ 4.100 \pm  0.028$ & $ 4.240 \pm  0.020$ & 2 &
    $ 4.530 \pm  0.030$ & $ 4.600 \pm  0.030$ & $ 4.640 \pm  0.030$ & A \\
HD064802 & $ 4.530 \pm  0.040$ & $ 5.270 \pm  0.029$ & $ 5.450 \pm  0.023$ & 2 &
    $ 5.857 \pm  0.021$ & $ 5.995 \pm  0.036$ & $ 5.984 \pm  0.018$ & A \\
HD074273 & $ 4.820 \pm  0.040$ & $ 5.720 \pm  0.029$ & $ 5.920 \pm  0.023$ & 3 &
    $ 6.322 \pm  0.023$ & $ 6.432 \pm  0.034$ & $ 6.458 \pm  0.018$ & A \\
HD165024 & $ 2.740 \pm  0.035$ & $ 3.580 \pm  0.028$ & $ 3.660 \pm  0.020$ & 2 &
    $ 3.830 \pm  0.030$ & $ 3.880 \pm  0.030$ & $ 3.910 \pm  0.030$ & A \\
HD188209 & $ 4.580 \pm  0.035$ & $ 5.550 \pm  0.028$ & $ 5.620 \pm  0.020$ & 2 &
    $ 5.724 \pm  0.026$ & $ 5.834 \pm  0.023$ & $ 5.815 \pm  0.016$ & A \\
HD195986 & $ 5.900 \pm  0.026$ & $ 6.470 \pm  0.021$ & $ 6.580 \pm  0.019$ & 3 &
    $ 6.749 \pm  0.020$ & $ 6.859 \pm  0.033$ & $ 6.868 \pm  0.018$ & A \\
HD204172 & $ 4.920 \pm  0.030$ & $ 5.850 \pm  0.024$ & $ 5.930 \pm  0.020$ & 4 &
    $ 6.085 \pm  0.021$ & $ 6.080 \pm  0.046$ & $ 6.141 \pm  0.018$ & A \\
HD214680 & $ 3.640 \pm  0.035$ & $ 4.680 \pm  0.028$ & $ 4.880 \pm  0.020$ & 2 &
    $ 5.303 \pm  0.037$ & $ 5.436 \pm  0.016$ & $ 5.498 \pm  0.029$ & A \\ \hline
\multicolumn{9}{c}{Reddened Stars} \\ \hline
BD+63D1964 & $ 8.790 \pm  0.033$ & $ 9.170 \pm  0.014$ & $ 8.460 \pm  0.010$ & 5 &
    $ 6.899 \pm  0.020$ & $ 6.715 \pm  0.031$ & $ 6.593 \pm  0.016$ & A \\
HD014956 & $ 7.620 \pm  0.035$ & $ 7.910 \pm  0.028$ & $ 7.190 \pm  0.020$ & 6 &
     $ 5.680 \pm  0.019$ & $ 5.466 \pm  0.023$ & $ 5.348 \pm  0.021$ & A \\
HD029309 & $ 7.110 \pm  0.017$ & $ 7.420 \pm  0.014$ & $ 7.100 \pm  0.010$ & 7 &
     $ 6.122 \pm  0.020$ & $ 6.041 \pm  0.024$ & $ 5.926 \pm  0.017$ & A \\
HD029647 & $ 9.690 \pm  0.039$ & $ 9.220 \pm  0.025$ & $ 8.310 \pm  0.017$ & 8 &
     $ 5.960 \pm  0.030$ & $ 5.593 \pm  0.021$ & $ 5.363 \pm  0.020$ & A \\
HD034921 & $ 6.790 \pm  0.017$ & $ 7.650 \pm  0.014$ & $ 7.510 \pm  0.010$ & 5 &
     $ 6.696 \pm  0.019$ & $ 6.507 \pm  0.016$ & $ 6.623 \pm  0.017$ & A \\
HD096042 & $ 7.620 \pm  0.049$ & $ 8.410 \pm  0.028$ & $ 8.230 \pm  0.020$ & 9 &
     $ 8.126 \pm  0.019$ & $ 8.186 \pm  0.057$ & $ 8.035 \pm  0.024$ & A \\
HD112272 & $ 7.930 \pm  0.040$ & $ 8.140 \pm  0.029$ & $ 7.340 \pm  0.023$ & 3 &
     $ 5.573 \pm  0.024$ & $ 5.363 \pm  0.026$ & $ 5.176 \pm  0.020$ & A \\
HD147701 & $ 8.823 \pm  0.017$ & $ 8.904 \pm  0.014$ & $ 8.360 \pm  0.010$ & 10 &
     $ 6.668 \pm  0.020$ & $ 6.378 \pm  0.040$ & $ 6.186 \pm  0.018$ & A \\
HD147889 & $ 8.580 \pm  0.017$ & $ 8.750 \pm  0.014$ & $ 7.920 \pm  0.010$ & 11 &
     $ 5.344 \pm  0.020$ & $ 4.938 \pm  0.076$ & $ 4.582 \pm  0.017$ & A \\
HD147933 & $ 4.300 \pm  0.035$ & $ 4.850 \pm  0.028$ & $ 4.630 \pm  0.020$ & 2 &
     $ 3.720 \pm  0.030$ & $ 3.540 \pm  0.030$ & $ 3.480 \pm  0.030$ & A \\
HD149038 & $ 4.200 \pm  0.035$ & $ 4.990 \pm  0.028$ & $ 4.890 \pm  0.020$ & 1 &
     $ 4.720 \pm  0.020$ & $ 4.700 \pm  0.010$ & $ 4.680 \pm  0.020$ & A \\
HD149404 & $ 4.810 \pm  0.014$ & $ 5.850 \pm  0.014$ & $ 5.460 \pm  0.010$ & 12 &
     $ 4.606 \pm  0.290$ & $ 4.387 \pm  0.232$ & $ 4.191 \pm  0.036$ & A \\
HD152408 & $ 5.170 \pm  0.035$ & $ 5.930 \pm  0.028$ & $ 5.770 \pm  0.020$ & 13 &
     $ 5.210 \pm  0.030$ & $ 5.020 \pm  0.030$ & $ 4.920 \pm  0.030$ & B \\
HD166734 & $ 9.390 \pm  0.017$ & $ 9.510 \pm  0.014$ & $ 8.420 \pm  0.010$ & 5 &
     $ 5.881 \pm  0.020$ & $ 5.517 \pm  0.020$ & $ 5.316 \pm  0.018$ & A \\
HD169454 & $ 7.300 \pm  0.042$ & $ 7.530 \pm  0.042$ & $ 6.630 \pm  0.030$ & 1 &
     $ 4.420 \pm  0.030$ & $ 4.090 \pm  0.030$ & $ 3.850 \pm  0.030$ & C \\
HD192660 & $ 7.810 \pm  0.010$ & $ 8.170 \pm  0.010$ & $ 7.380 \pm  0.030$ & 1 &
     $ 6.151 \pm  0.018$ & $ 6.004 \pm  0.017$ & $ 5.909 \pm  0.016$ & A \\
HD197702 & $ 7.390 \pm  0.017$ & $ 8.080 \pm  0.014$ & $ 7.900 \pm  0.010$ & 7 &
     $ 7.186 \pm  0.027$ & $ 7.084 \pm  0.021$ & $ 6.940 \pm  0.018$ & A \\
HD204827 & $ 8.610 \pm  0.017$ & $ 8.760 \pm  0.014$ & $ 7.950 \pm  0.010$ & 5 &
     $ 6.472 \pm  0.020$ & $ 6.356 \pm  0.021$ & $ 6.319 \pm  0.016$ & A \\
HD206773 & $ 6.170 \pm  0.064$ & $ 7.010 \pm  0.023$ & $ 6.790 \pm  0.016$ & 14 &
     $ 6.071 \pm  0.024$ & $ 5.910 \pm  0.030$ & $ 5.567 \pm  0.017$ & A \\
HD229059 & $10.660 \pm  0.035$ & $10.230 \pm  0.028$ & $ 8.700 \pm  0.020$ & 15 &
     $ 5.551 \pm  0.032$ & $ 5.083 \pm  0.018$ & $ 4.825 \pm  0.018$ & A \\
HD229238 & $ 9.720 \pm  0.017$ & $ 9.780 \pm  0.014$ & $ 8.880 \pm  0.010$ & 5 &
     $ 7.015 \pm  0.023$ & $ 6.822 \pm  0.017$ & $ 6.717 \pm  0.016$ & A \\
HD281159 & $ 9.220 \pm  0.010$ & $ 9.210 \pm  0.014$ & $ 8.530 \pm  0.010$ & 5 &
     $ 6.789 \pm  0.019$ & $ 6.640 \pm  0.021$ & $ 6.515 \pm  0.017$ & A \\
HD283809 & $12.630 \pm  0.039$ & $12.140 \pm  0.034$ & $10.720 \pm  0.017$ & 8 &
     $ 6.994 \pm  0.030$ & $ 6.461 \pm  0.030$ & $ 6.169 \pm  0.030$ & A \\
CYGOB2-1 & $12.790 \pm  0.035$ & $12.480 \pm  0.028$ & $11.060 \pm  0.020$ & 16 &
     $ 8.050 \pm  0.020$ & $ 7.610 \pm  0.020$ & $ 7.320 \pm  0.020$ & A \\
CYGOB2-2 & $11.890 \pm  0.035$ & $11.760 \pm  0.028$ & $10.610 \pm  0.020$ & 15 &
     $ 8.140 \pm  0.020$ & $ 7.820 \pm  0.020$ & $ 7.540 \pm  0.020$ & A \\
CYGOB2-5 & $11.390 \pm  0.035$ & $10.830 \pm  0.028$ & $ 9.150 \pm  0.020$ & 15 &
     $ 5.340 \pm  0.020$ & $ 5.060 \pm  0.020$ & $ 4.480 \pm  0.020$ & A \\
CYGOB2-8A & $10.480 \pm  0.035$ & $10.320 \pm  0.028$ & $ 9.050 \pm  0.020$ & 5 &
     $ 6.123 \pm  0.020$ & $ 5.721 \pm  0.017$ & $ 5.503 \pm  0.017$ & A
\enddata
\tablerefs{(1) \citet{1983ApJS...52....7F};
(2) \citet{1966CoLPL...4...99J}; (3) \citet{1983ApJS...51..321S}; (4) \citet{1977AJ.....82..431L}; (5) \citet{1956ApJS....2..389H}; (6) \citet{1967BOTT....4..149M}; (7) \citet{1974PASP...86..795G}; (8) \citet{1980SvAL....6..397S}; (9) \citet{1969MNRAS.143..273F}; (10) \citet{1968ApJS...15..459G}; (11) \citet{1973MNSSA..32..117C}; (12) \citet{1979AJ.....84.1713F}; (13) \citet{1969ApJ...156..609S}; (14) \citet{1976PASP...88..865G}; (15) \citet{1978AaAS...34....1N}; (16) \citet{1956ApJ...124..367H};
(A) \citet{2006AJ....131.1163S}; (B) \citet{1984AaA...132..151L}; (C) \citet{1976MNRAS.177..625W} }
\end{deluxetable*}

We selected the sample to probe sightlines with existing or planned UV extinction curve measurements.
The sightlines in the sample (listed in Table~\ref{tab_phot}) span the known range of $R(V)$ values and as large a range of $A(V)$ values as possible.
This sample allows for direct comparison of the features and the overall level of the UV and MIR extinction curves.
Our sample included both reddened and comparison stars to allow the standard pair method to be used to measure the dust extinction.

The majority of the IRS, IRAC, \& MIPS observations for this work were taken as part of two Spitzer programs (PIDs: 20146 \& 50043) specifically dedicated to studying the diffuse interstellar medium IR dust extinction and its connection to the UV dust extinction.
The stars, HD~14422 and NGC2024~1, were observed as part of the 1st program, but are not included in this study as the Spitzer photometry and spectroscopy were not consistent with the NIR and optical photometry indicating an issue with the observations.
Archival Spitzer observations were used where they existed to complete the coverage.

\subsection{Spitzer Photometry}
\label{sec_spit_phot}

The IRAC photometry was taken in subarray mode for all the stars as they are too bright for IRAC full frame observations.
Each star was observed in all four IRAC bands ($\lambda_o =$ 3.5, 4.5, 5.7, and 7.8~\micron) with each band observation being composed of 64 or 256 individual images.
Aperture photometry was done on each of the individual images using a 3 pixel radius aperture and a sky annulus with radii of 10 and 20 pixels.
The flux and measurement uncertainty in each band were determined from a sigma clipped average and standard deviation of the good individual image measurements.
The fluxes were corrected to the IRAC standard aperture of 10 pixels using the appropriate aperture corrections \citep{Hora08}, calibration factors \citep{Reach05}, and small updates to the calibration factors from \citet{Bohlin11}.
A 2\% absolute flux uncertainty was added in quadrature to the measurement uncertainty to determine the final uncertainty for each band \citep{Reach05, Bohlin11}.

The IRS blue peakup photometry ($\sim$15~\micron) was taken using the standard photometry mode with one cycle of dithered 6~s exposures resulting in 5 images per target.
Aperture photometry was done on each observation mosaic with a 12 pixel radius aperture and a sky annulus with radii of 18 and 28 pixels.
The measurement uncertainty in each band was determined using the noise measured in the sky annulus.
The fluxes were corrected using an aperture correction of 1.077 and a color correction of 1.038 taken from the IRS Instrument Handbook.
A 2\% absolute flux uncertainty was added in quadrature to the measurement uncertainty to determine the final uncertainty for each band (assumed similar to IRAC and MIPS 24~\micron).
For 4 stars, IRS blue peakup imaging was not possible as the stars were too bright.

The MIPS 24~\micron\ photometry was taken using the standard photometry mode with one cycle of dithered 3~s exposures.
The data were reduced using the MIPS Data Analysis Tool \citep{Gordon05DAT}.
Additional processing steps specific to photometry mode data were performed to remove residual instrumental signatures.
These additional steps included applying scan mirror dependent flat fields and a final scan mirror independent flat field.
Point-source fitting (PSF) photometry was done on the mosaic of the exposures for each source to provide high quality measurements even in the presence of nearby sources.
The PSF used and conversion to physical units are from \citet{Engelbracht07MIPS24}.
A 2\% absolute flux uncertainty was added in quadrature to the measurement uncertainty to determine the final uncertainty \citep{Engelbracht07MIPS24}.
Measurements of the 24~\micron\ photometry were not possible for HD~96042 and Cyg~OB2~8a due to nearby extended emission.

The MIPS 24~\micron\ images are shown in Appendix~\ref{spitzer_images} with the IRS spectroscopic apertures overlaid.
These images clearly show that the majority of our targets are well isolated point sources at the MIPS 24~\micron\ spatial resolution.
The images are useful for diagnosing potential issues with observations of particular stars as the MIPS 24~\micron\ images have a 5\arcmin\ field-of-view.

The IRAC, IRS blue peakup (IRSB), and MIPS 24~\micron\ photometry and wavelength coverage of the IRS spectroscopy are given for both comparison and reddened stars in Table~\ref{tab_phot}.
The fluxes were converted to Vega magnitudes using the conversion factors 65.0, 26.6, 10.2, 3.04, 0.194, and 0.0383 $\times 10^{-13}$~ergs~s$^{-1}$~cm$^{-2}$~\AA$^{-1}$ for IRAC1, IRAC2, IRAC3, IRAC4, IRSB, and MIPS24 bands, respectively (IRAC \& MIPS Handbooks; R.\ Bohlin, priv. comm.).

\subsection{IRS Spectroscopy}

\begin{figure*}[tbp]
\epsscale{1.1}
\plotone{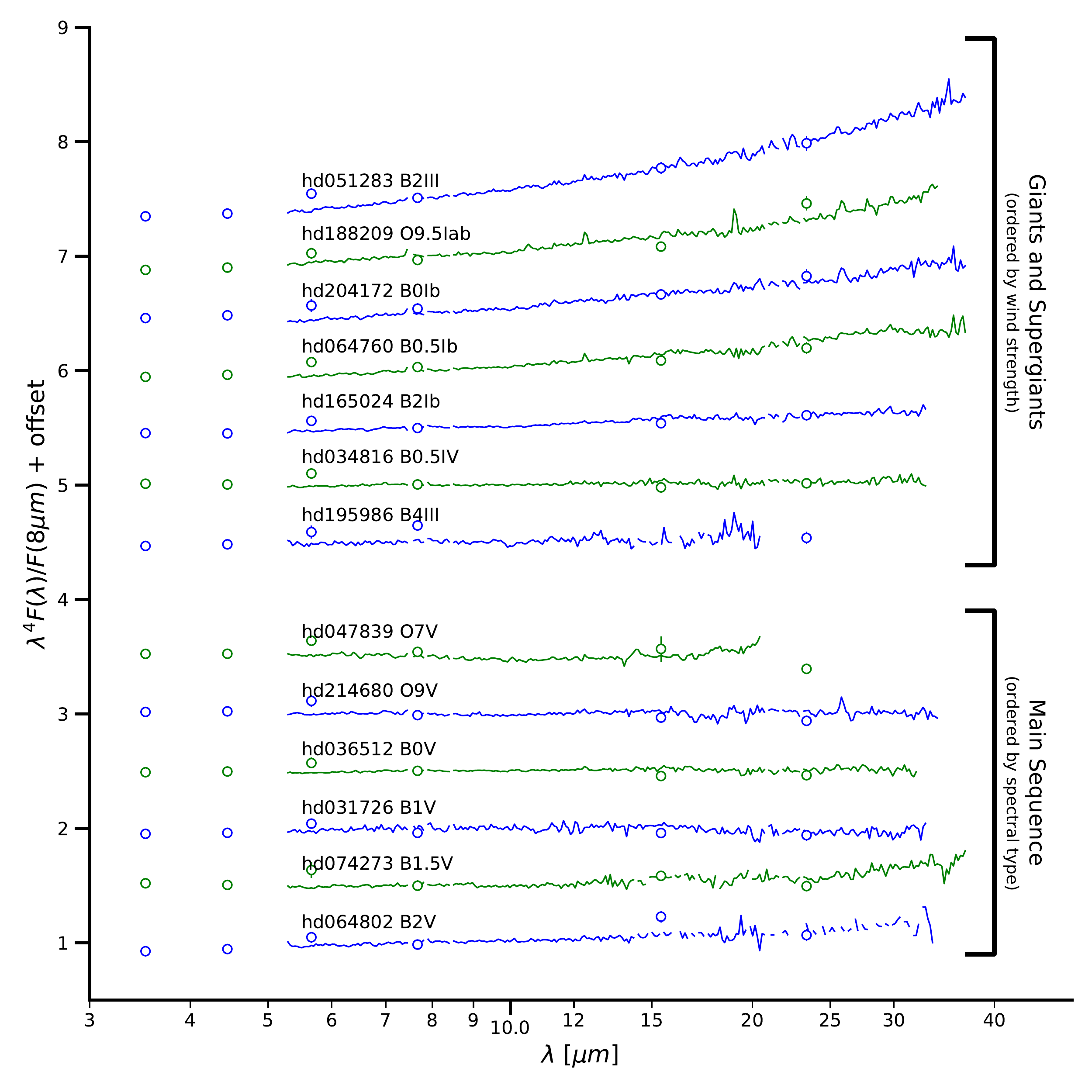}
\caption{The infrared SEDs of the comparison stars are plotted. The IRS spectra are shown as solid lines and the IRAC, IRS blue peakup, and MIPS~24~\micron\ photometry is shown as circles.
All the data have been multiplied by $\lambda^4$ to remove the expected shape of an unreddened hot star (the Rayleigh-Jeans tail of a black body).
Each SED has been normalized at 8~\micron\ and offset by a constant value in log space.
\label{fig_ir_seds_standards}}
\end{figure*}

\begin{figure*}[tbp]
\epsscale{1.1}
\plotone{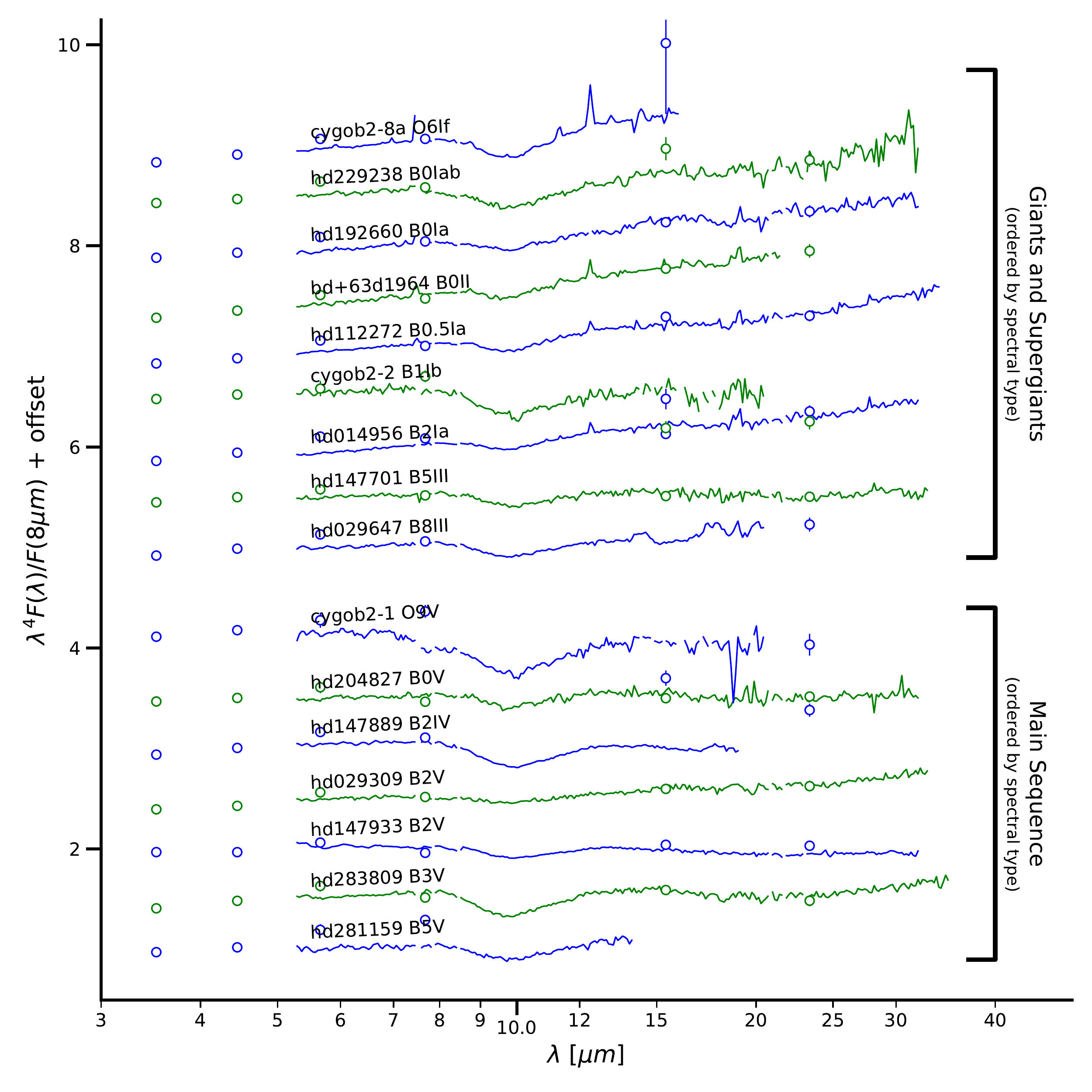}
\caption{The infrared SEDs of the reddened stars without strong winds are plotted.
The IRS spectra are shown as solid lines and the IRAC, IRS blue peakup, and MIPS~24~\micron\ photometry is shown as open circles.
All the data have been multiplied by $\lambda^4$ to remove the expected shape of an unreddened hot star (the Rayleigh-Jeans tail of a black body).
Each SED has been normalized at 8~\micron\ and offset by a constant value in log space.
\label{fig_ir_seds_reddened}  }
\end{figure*}

\begin{figure*}[tbp]
\epsscale{1.1}
\plotone{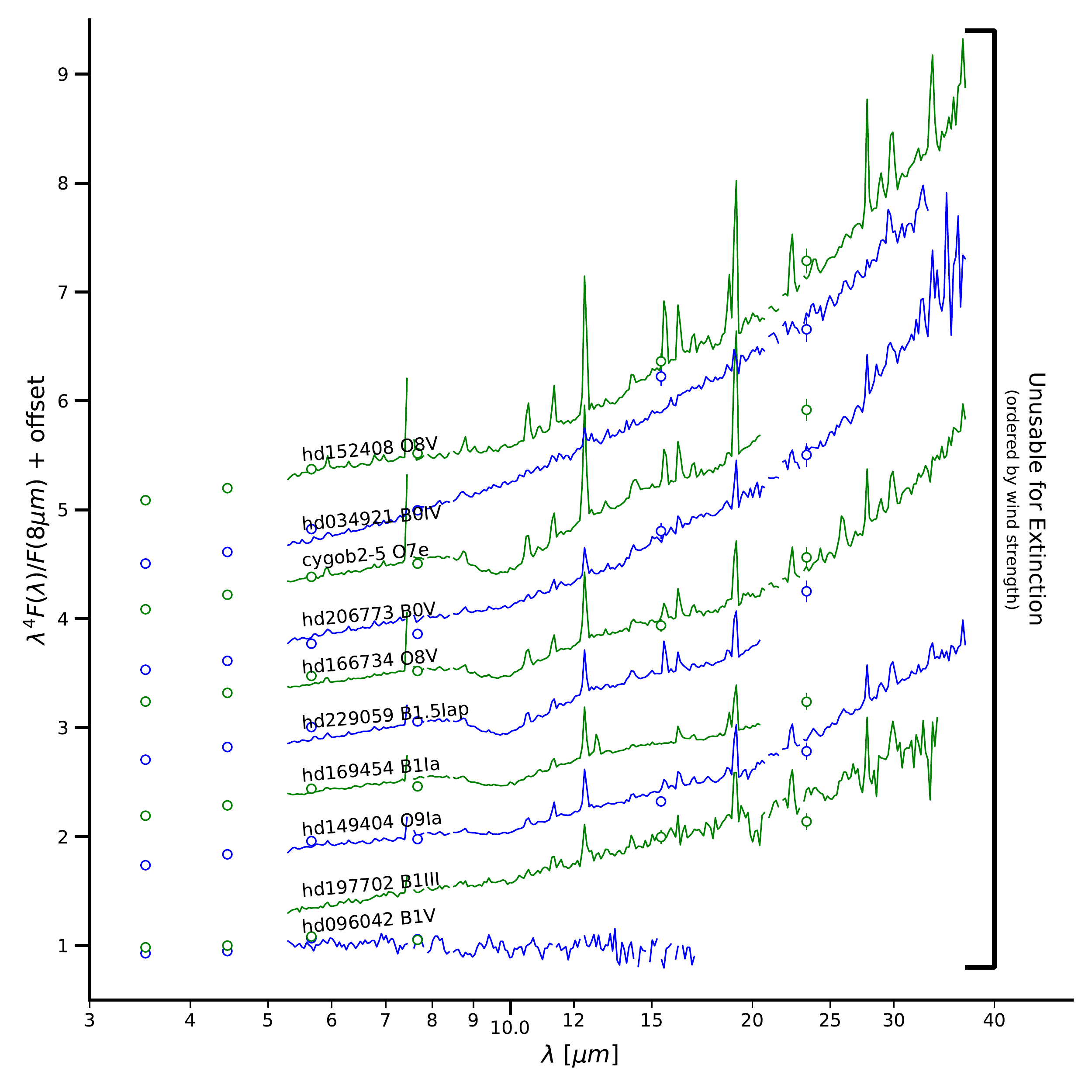}
\caption{The infrared SEDs of the reddened stars with strong winds and the star with a poor IRS spectrum are plotted.  The IRS spectra are shown as solid lines and the IRAC, IRS blue peakup, and MIPS~24~\micron\ photometry is shown as open circles.
All the data have been multiplied by $\lambda^4$ to remove the  expected shape of an unreddened hot star (the Rayleigh-Jeans tail of a black body).
Each SED has been normalized at 8~\micron\ and offset by a constant value in log space.
\label{fig_ir_seds_reddened_windy}  }
\end{figure*}

There is IRS spectroscopy for all the stars, at least for  5--21~\micron.
For the stars observed as part of the PID 20146 program, the IRS spectroscopy extends to $\sim$38~\micron\ (see Table~\ref{tab_phot}).
The IRS spectral apertures overlaid on the MIPS~24~\micron\ images are shown in Appendix~\ref{spitzer_images}.
The planned spectroscopic observations for HD149038 unfortunately observed a nearby, brighter source, hence there is no IRS spectrum for this source.

The IRS spectra were extracted for most stars using two independent methods.
This approach was adopted to test the spectral fidelity, since it is critical to obtain the best quality spectral extractions given that the measurement of the continuum extinction is sensitive to the continuum accuracy.
The first extraction method used custom routines and the second used SMART \citep{Higdon04}.
Details of both reduction methods are given below.
The extractions were found to be very similar with the best varying between the custom and SMART extractions.
We visually examined all the stars with both extractions and used the one that has the best S/N and fewest artifacts as indicated in the ``Extraction'' column of Table~\ref{tab_phot}.
As part of the visual inspections, we determined the maximum usable wavelength to avoid including clear artifacts likely due to the low sensitivity of IRS at long wavelengths.
The usable wavelength range is given in the ``IRS'' column of Table~\ref{tab_phot}.
In addition to our manual extractions, fully reduced spectra were downloaded from the CASSIS \citep{Lebouteiller11CASSIS} database\footnote{\url{https://cassis.sirtf.com}} for all our objects.
The automated spectral extractions were compared to our custom extractions; in all cases, the custom and SMART extractions had higher S/N, fewer artifacts, and similar spectral slopes.

For the custom extraction, the data reduction process started from the \emph{droopres} intermediate data product processed through the Spitzer Science Center pipeline version S18.18.0.
For the spectral extraction and flux calibration we used the data reduction packages developed for the c2d and feps legacy programs \citep{Lahuis2003, Bouwman2008, Carpenter2008}.
For those observations with a uniform background emission in the immediate surroundings of the target stars, the background emission has been subtracted using the associated pairs of the spectral images from the two nod positions, also eliminating stray light contamination and anomalous dark currents.
For all others we subtracted the background by fitting a low order polynomial to the observed background near the target star and used this to subtract the background emission.
Pixels flagged by the data pipeline as being ``bad'' were replaced with a value interpolated in the dispersion direction from an elongated 8~pixel perimeter surrounding the flagged pixel.
The spectra were extracted using a 6~pixel and 5~pixel fixed-width aperture in the spatial dimension for the short-low (SL) and the long-low (LL) modules, respectively.
The low-level fringing at wavelengths $>20\mu$m due to known filter delamination was removed using the irsfringe package \citep{Lahuis2003}.
The spectra were calibrated with a position-dependent spectral response function derived from IRS spectra and MARCS stellar models for a suite of calibrators provided by the Spitzer Science Center.
To remove any effect of pointing offsets in the SL module data, we matched orders based on the point spread function of the IRS.

The SMART extractions used SMART version 8.2.1 with Advanced Optimal extraction \citep{Higdon04SMART, Lebouteiller10SMARTAdOpt}.
Raw data were imported into a SMART project, cleaned using the IRSCLEAN routine, and then all individual, cleaned, images for each exposure id were combined into a single exposure for each module (SL2, SL1, LL2 and sometimes LL1) and nod position.
For each module, the two nod positions were subtracted, resulting in 2 individual 2 dimensional spectra.
One dimensional spectra were extracted from each of the 2 nod difference positions using the manual optimal extraction option.
While nod subtraction removes the bulk of the background (as well as rogue pixels), for complex backgrounds some residuals can remain, especially at longer wavelengths.
For each object, module, and nod position, different background polynomial orders from ``none'' to 2 were examined across the full wavelength range of the extraction to minimize residuals.
In general, the maximum order of any additional background subtraction at this stage was linear.
After optimal extraction, the individual nod positions were trimmed and combined into a single spectrum for each module.
The spectral segments from the separate modules were combined into a single full spectrum using the overlap regions. The scaling between the SL1 and SL2 modules was minimal, less than 2\% and, in most cases, no offsets were required.
To match the LL2/LL1 to SL modules in the overlap region, larger offsets in the LL modules were required, generally $\le$5\%.
In 3 cases (HD~29647, HD~147889, and Cyg~OB2~1), the LL2 module required a scaling of $\sim$20\%.
After matching the modules internally, the final full spectrum was scaled to the wide band photometric fluxes using synthetic photometry with the appropriate filter response curve.

We checked the overall level and spectral shape of the IRS spectra using the IRAC, IRS 15~\micron, and MIPS 24~\micron\ photometry (see \S\ref{sec_spit_phot}).
The fixed IRS slit widths combined with a diffraction limited telescope can result in errors in the overall level and spectral slope due to small errors in the centering of the source in the slits.
For the majority of our stars, no corrections were needed.
For a small subset, small additional slit-loss corrections were needed to match the photometry in overall level (HD036512, HD096042, HD112272, HD152408, HD188209, HD229238, CygOB2-5) and in overall level and spectral slope (HD064802, HD204172).
These corrections were based on comparing the photometry to synthetic photometry created using the spectra and the band response functions and visually examining the spectra and photometry.
We found that the visual examination in plots where the spectra and photometric fluxes had been multiplied to remove the Rayleigh-Jeans slope (e.g., $F(\lambda)\lambda^4$) was particularly useful in identifying the corrections needed.

The IRS spectra and IRAC, IRSB, and MIPS photometry are plotted in Fig.~\ref{fig_ir_seds_standards}, \ref{fig_ir_seds_reddened}, and \ref{fig_ir_seds_reddened_windy}.
The IRS spectra are available electronically \citep{zenododata}.
The spectra are plotted multiplied by $\lambda^4$ to remove the strongly decreasing Rayleigh-Jeans tail expected for stars in this wavelength range.
Plotting MIR spectra this way means that stars with a bare stellar atmosphere alone should be approximately horizontal.
Stars with circumstellar shells produced by strong stellar winds or rapid rotation typically exhibit emission lines and/or a marked increase in infrared continuum flux toward longer wavelengths.
Such stars introduce additional uncertainty in measuring extinction using the standard pair method as a suitable comparison with the same wind emission is needed.

For the comparison stars, Fig.~\ref{fig_ir_seds_standards} illustrates that the main sequence stars have bare stellar atmosphere spectra, very close to Rayleigh-Jeans throughout the MIR.
The giants and supergiant comparison stars display a range from bare stellar atmospheres (e.g., HD034816) to stars with the clear wind signature of slopes that are redder than Rayleigh-Jeans (e.g., HD051283).
Some of the spectra showing indications of weak winds in the continuum also show line emission (e.g., HD188209).

\begin{figure}[tbp]
\epsscale{1.2}
\plotone{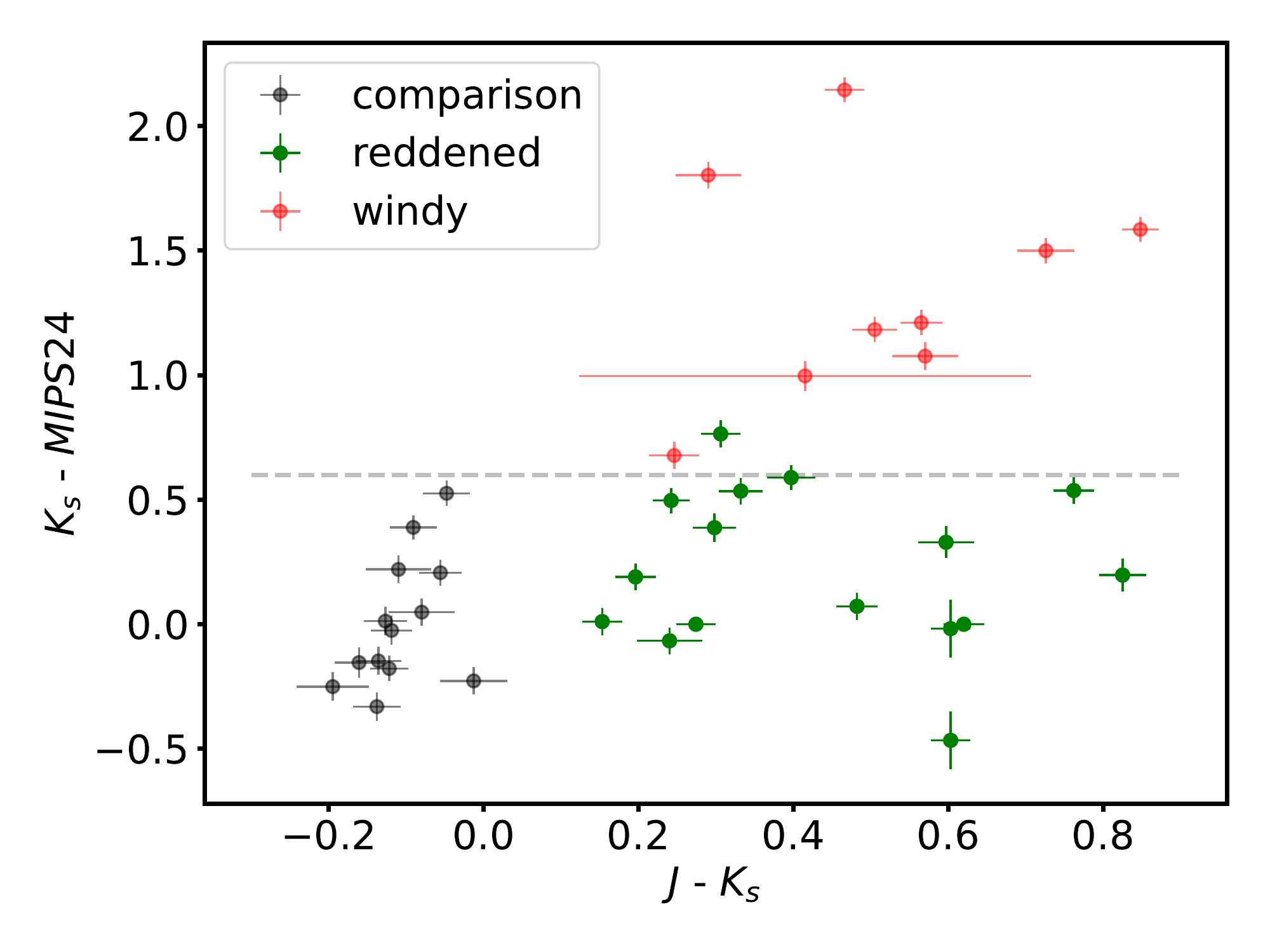}
\caption{The $J-K_s$ color is plotted versus the $K_s - \mathrm{MIPS24}$ color.
The $J - K_s$ color probes dust reddening and, to a lesser extent, the stellar temperature.
The $K_s - \mathrm{MIPS24}$ color probes the presence of a significant stellar wind.
The horizontal dash line at $K_s - \mathrm{MIPS24} = 0.6$~mag gives a reasonable dividing line between sources without and with winds.
\label{fig_nir_mir_phot}  }
\end{figure}

The spectra of the reddened stars can be split into those without strong wind indicators (Fig.~\ref{fig_ir_seds_reddened}) and those with either strong continuum and emission line wind indicators or very poor IRS spectra (Fig.~\ref{fig_ir_seds_reddened_windy}).
The signature of winds clearly seen in the IRS spectra can also be seen in a $J - K_s$ versus $K_s - \mathrm{MIPS24}$ color-color plot (Fig.~\ref{fig_nir_mir_phot}).
The $J - K_s$ color gives a measure of the reddening and the $K_s - \mathrm{MIPS24}$ color shows a wind signature for values greater than $\sim$0.6 mag.
The equivalent $K_s - \mathrm{IRAC4}$ color cut is $\sim$0.4 mag.
All the reddened main sequence stars with good IRS spectra are in the non-winds category and show overall Rayleigh-Jeans spectra with clear signature of the silicate extinction feature at 10~\micron.
The giant and supergiant stars in the non-windy category also show the signature of silicate extinction, but have slightly redder than Rayleigh-Jeans spectra.

The windy category of reddened stars clearly show the signature of winds in their continuum shape and presence of emission lines.
Most also show the silicate extinction feature at 10~\micron.
The lack of comparison stars with appropriately strong winds means that these stars cannot be used to calculate extinction curves in this study.

\subsection{Other Data}

The optical and NIR photometry for our sample stars is given in Table~\ref{tab_ophot}.
The ultraviolet spectroscopy was taken from the {\em IUE} and {\em HST} archives except for Cyg~OB2~1,  Cyg~OB2~2, Cyg~OB2~5, and Cyg~OB2~8A which do not have the appropriate UV spectroscopy available.
{\em IUE} and {\em HST} ultraviolet spectroscopy is often used to measure extinction along sightlines \citep[e.g.,][]{Fitzpatrick90, Valencic04, Clayton03UVDIBS}.

\section{Extinction Curves}
\label{sec_curves}

\begin{figure*}[tbp]
\epsscale{1.0}
\plotone{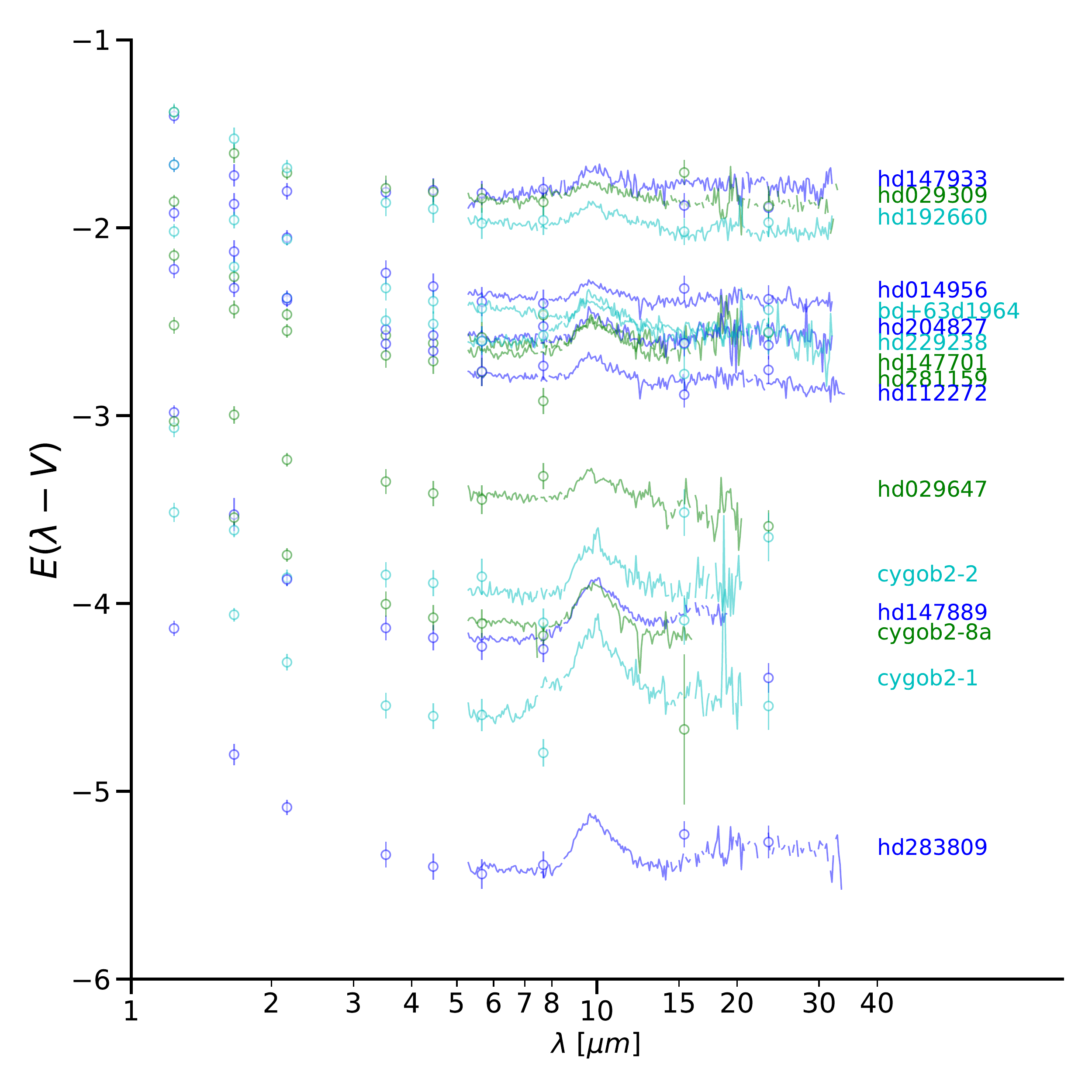}
\caption{The \elv\ extinction curves are given for all the sightlines where extinction curves can be measured.
The photometric extinction measurements are shown as open circles and the spectroscopic extinction measurements as solid lines.
The range of $A(V)$ values can be easily estimated as \elv\ $\rightarrow -A(V)$ as $\lambda \rightarrow \infty$ and is $\sim$1.5 to $\sim$5.5~mag. The varying level of the extinction curves results from different total columns of dust along each sightline; those with the lowest dust columns appear at the top and the highest appear at the bottom of the figure.
\label{fig_ir_elv}}
\end{figure*}

The extinction curves were computed using the pair method \citep{Stecher65, Massa83}.
The core of the pair method is the flux ratio between the reddened and comparison stars with an ``identical'' intrinsic spectrum.
The flux ratio is referenced to its value at a reference wavelength to remove the distance dependence from the extinction measurement \citep{Gordon09FUSE}.
The resulting measurement is expressed as a color excess in units of magnitudes.
We use the V band as the wavelength reference and this makes $E(\lambda - V)$ our extinction measurement.
Uncertainties on the extinction measurements were calculated based on the flux uncertainties.
For the IRS spectra, we imposed a maximum signal-to-noise of 100 to conservatively account for the known overestimation of their signal-to-noise.

\begin{deluxetable}{lclc}
\tablewidth{0pt}
\tablecaption{Sightline Properties \label{tab_red_sample}}
\tablehead{\colhead{Name} & \colhead{SpType} & \colhead{comparison} &
\colhead{Type} }
\startdata
\multicolumn{3}{c}{Main Sequence} \\ \hline
CygOB2-1 & O9V & HD214680 & diffuse \\
HD204827 & B0V & HD036512 & diffuse \\
HD147889 & B2IV & HD064802 & diffuse \\
HD029309 & B2V & HD064802 & diffuse \\
HD147933 & B2V & HD031726 & diffuse \\
HD283809 & B3V & HD064802 & dense \\
HD281159 & B5V & HD064802 & diffuse \\ \hline
\multicolumn{3}{c}{Giants and Supergiants} \\ \hline
CygOB2-8A & O6If & HD188209 & diffuse \\
HD229238 & B0Iab & HD204172 & diffuse \\
HD192660 & B0Ia & HD204172 & diffuse \\
BD+63D1964 & B0II & HD188209 & diffuse \\
HD112272 & B0.5Ia & HD204172 & diffuse \\
CygOB2-2 & B1b & HD214680 & diffuse \\
HD014956 & B2Ia & HD204172 & diffuse \\
HD147701 & B5III & HD195986 & diffuse \\
HD029647 & B8II & HD195986 & dense
\enddata
\end{deluxetable}

The sightlines for which extinction measurements can be made are given in Table~\ref{tab_red_sample} including the spectral type of the reddened stars and the name of the comparison star used.
The extinction curves were calculated using the `measure\_extinction' package \citep{measureextinction}, and are available electronically \citep{zenododata}.
The diffuse versus dense designation is discussed in \S\ref{sec_results}.
The comparison stars were matched based on spectral type \citep{Wenger00} with the goal of being the closest match given the possible comparison stars.
For the giants and supergiants, we found  matching the strength of the continuum wind signature is also important for producing reasonable extinction curves.
Note that we were only able to measure extinction curves for 16 reddened stars as the spectra of 11 stars were complicated by strong winds, non-standard stellar continuum, or the wrong star was observed.
This is in contrast with the UV, where all the stars except for the 4 without UV spectra have measured UV to NIR extinction curves.
Winds can strongly affect the MIR continuum, whereas in the UV they mainly affect narrow spectral regions around wind sensitive lines (e.g., \ion{C}{4}).

The \elv\ extinction curves for the IR wavelength region are given in Figure~\ref{fig_ir_elv}.
All of these clearly show the 10~\micron\ silicate extinction feature with a subset also showing the weaker 20~\micron\ silicate extinction feature.
There are additional smaller features in individual curves, but their central wavelengths and widths vary between the different sightlines.
These additional ``features'' are most likely due to residual instrumental effects (see Figs.~\ref{fig_ir_seds_standards} \& \ref{fig_ir_seds_reddened}).
The varying level of the extinction curves results from different total columns of dust along each sightline; those with the lowest dust columns appear at the top and the highest appear at the bottom of the figure.

\subsection{Fitting and Total Column Normalization}
\label{sec_normalization}

Measuring the total column along each sightline allows for the dust along different sightlines to be compared per unit amount of dust.
The two most common dust column measurements are \ebv\ (e.g., color excess) and $A(V)$.
In cases where it is not possible to measure \ebv\ due to these wavelengths not being observed, then an alternative color excess can be used (e.g., $E(J-K)$).
The drawback of using any color excess for normalization is that it is the difference of the extinction at two wavelengths, an indirect measurement of the total column.

$A(V)$ is a more direct probe of the total dust column than \ebv, but it requires an additional step to measure.
Like \ebv, if the V band was not observed another wavelength can be used (e.g, $A(K)$).
The additional step needed is to extrapolate the \elv\ measurement to derive $A(V)$ as $A(\lambda = \infty) = 0.0$.
The accuracy of this extrapolation is related to the longest wavelength used as dust grains are most efficient at generating differential extinction at wavelengths similar to their sizes.
The extrapolation method is based on the differential extinction measurement (e.g., $E(\lambda - V)$) and is insensitive to extinction components that are not wavelength dependent.
For example, the extinction from grains larger than the longest wavelength measured would not be included in such extrapolation measurements of $A(V)$.
While our observations extend to 30+~\micron, the fact that grains with a range of sizes contribute to the 10 and 20~\micron\ silicate features means that it is the observations up to $\sim$6~\micron\ that place the strongest constraints on $A(V)$.
Fortunately, most dust grains are expected to be significantly smaller than 6~\micron\ \citep{Weingartner01, Zubko04, Jones13} indicating that our measurements of $A(V)$ should be quite robust.

A common extrapolation method is based on assuming the NIR extinction behaves as $A(\lambda) \propto \lambda^{-\alpha}$ \citep{Martin90}.
However, our MIR extinction curves show clear structure beyond a simple power law, specifically the strong 10~\micron\ feature, requiring a more complicated functional form.
We use a combination of a power law and two modified Drude functions to fit the measured NIR/MIR extinction curves to derive $A(V)$ values and the strengths of the 10 and 20~\micron\ silicate features.
We investigated using other functional forms for the fitting, including one based on that of \citet{Pei92} and one based on polynomials plus two unmodified Drude profiles.
These other functional forms produced larger residuals and were less stable than the one we adopted.
The full functional form fit to the measured extinction curves is
\begin{equation}
\label{eq_mir_elv}
E(\lambda - V) = A(V) \left[ k(\lambda) - 1 \right] .
\end{equation}
where
\begin{eqnarray}
\label{eq_mir_ext}
k(\lambda) & = & \frac{A(\lambda)}{A(V)} \\
 & = & B \lambda^{-\alpha} + S_1 D_m(\lambda) + S_2 D_m(\lambda) \label{eq_func}
\end{eqnarray}
where $B$ is the amplitude of the power law component, $\alpha$ is the exponent of the power law, $S_i$ is the amplitude of each of the silicate features, and $D_m(\lambda)$ is a modified Drude profile.
A regular Drude profile is intrinsically asymmetric, but profiles with the appropriate central wavelength and width are not asymmetric enough to fit the silicate feature.
The modified Drude profile was inspired by the work of \citet{Stancik08} that introduced simple asymmetric profiles to model infrared absorption peaks based on Lorentzian and Gaussian functions with widths that vary continuously across the profiles.
The definition for the modified Drude is
\begin{eqnarray}
D_m(\lambda) & = & \frac{(\gamma/\lambda_o)^2}{(\lambda / \lambda_o - \lambda_o / \lambda)^2 + (\gamma / \lambda_o)^2}, \\
\gamma & = & \frac{2 \gamma_o}{1 + \exp [a (\lambda - \lambda_o)]},
\end{eqnarray}
$\lambda_o$ is the central wavelength, $\gamma_o$ is the unmodified width, and $a$ is the ``extra'' asymmetry parameter.

A Drude profile is often used to fit dust features including the 2175~\AA\ bump \citep{Fitzpatrick86}, the 10 and 20~\micron\ silicate features \citep{Pei92}, the aromatic/PAH emission features \citep{Smith07}, and the broad optical features \citep{Massa20}.
Consistent with the strong observed asymmetry in this feature \citep{Kemper04}, the modified Drude profile was found to fit the 10~\micron\ silicate feature better than the regular Drude profile.
We do not claim a physical origin for the modified Drude profile fits.
The goal of using modified Drude profiles is to have an analytic function with the fewest parameters that fits the NIR/MIR extinction curve.
Other formulations for fitting the silicate features have been explored in the literature both based on more complex analytic functions \citep[e.g.,][]{Shao18} and laboratory measurements of candidate materials \citep[e.g.,][]{Kemper04, Min07, Speck11}.

\begin{deluxetable}{cccl}
\tablewidth{0pt}
\tablecaption{Extinction Fit Parameters \label{tab_mir_param}}
\tablehead{\colhead{parameter} & \colhead{range} & \colhead{units} & \colhead{description} }
\startdata
\multicolumn{4}{c}{NIR+MIR} \\ \hline
$A(V)$ & 0--8 & mag & V absolute extinction \\
$B$ & 0--1 & $A(\lambda)/A(V)$ & power law amplitude \\
$\alpha$ & 0.5--5 &  & power law exponent \\
$S_1$ & 0.001--0.3 & $A(\lambda)/A(V)$ & 10~\micron\ silicate amplitude \\
$\lambda_{o1}$ & 8--12 & \micron & 10~\micron\ silicate center \\
$\gamma_{o1}$ & 1--10 & \micron & 10~\micron\ silicate width \\
$a_1$ & -2--2 & & 10~\micron\ silicate asymmetry \\
$S_2$ & 0.001--0.3 & $A(\lambda)/A(V)$ & 20~\micron\ silicate amplitude \\
$\lambda_{o2}$ & 16--24 & \micron & 20~\micron\ silicate center \\
$\gamma_{o2}$ & 5--20 & \micron & 20~\micron\ silicate width \\
$a_2$ & -2--2 & & 20~\micron\ silicate asymmetry \\ \hline
\multicolumn{4}{c}{UV} \\ \hline
C1 & -2--3 & $A(\lambda)/A(V)$ & linear term zero point \\
C2 & -0.1--1.0 & $A(\lambda)/A(V)$ & linear term slope \\
C3 & 0--2.5 & $A(\lambda)/A(V)$ & 2175~\AA\ bump amplitude \\
C4 & 0--1 & $A(\lambda)/A(V)$ & FUV rise amplitude \\
$x_o$ & 4.5--4.9 & $\micron^{-1}$ & 2175~\AA\ bump center \\
$\gamma$ & 0.6--1.5 & $\micron^{-1}$ & 2175~\AA\ bump width \\
\enddata
\end{deluxetable}

The full set of parameters fit to each extinction curve from 1 to 38~\micron\ are listed in Table~\ref{tab_mir_param} along with their fitting ranges.
The fitting ranges were initially chosen to encompass the expected range for each parameter based on previous work and then refined to encompass the full range seen in our sample.
The fitting was done using the `emcee' v3 MCMC sampling package \citep{Foreman-Mackey13} where the fitting ranges were implemented as flat priors.
In detail, Eq.~\ref{eq_func} was implemented as an astropy \citep{astropy:2013, astropy:2018} model (hereafter `G21') and this model was combined with the `AxAvToExv' model to implement Eq.~\ref{eq_mir_elv}.
Both models are provided as part of the `dust\_extinction' package \citep{dustextinction}.
For the entire sample, we found that our measurements were not sensitive to the 20~\micron\ silicate width and so we fixed $\gamma_{o,2} = 13$~\micron.
Approximately half of our sample only has spectroscopic data fully covering the 10~\micron\ silicate feature with partial or no coverage of the 20~\micron\ feature.
For this portion of the sample, we fixed the majority (or all) of the 20~\micron\ silicate feature parameters to reasonable values.
The MIR fit parameters are given in Tables~\ref{tab_mir_gen_ext_params} and \ref{tab_mir_sil_ext_params} with fixed parameters given without uncertainties.

\begin{figure}[tbp]
\epsscale{1.2}
\plotone{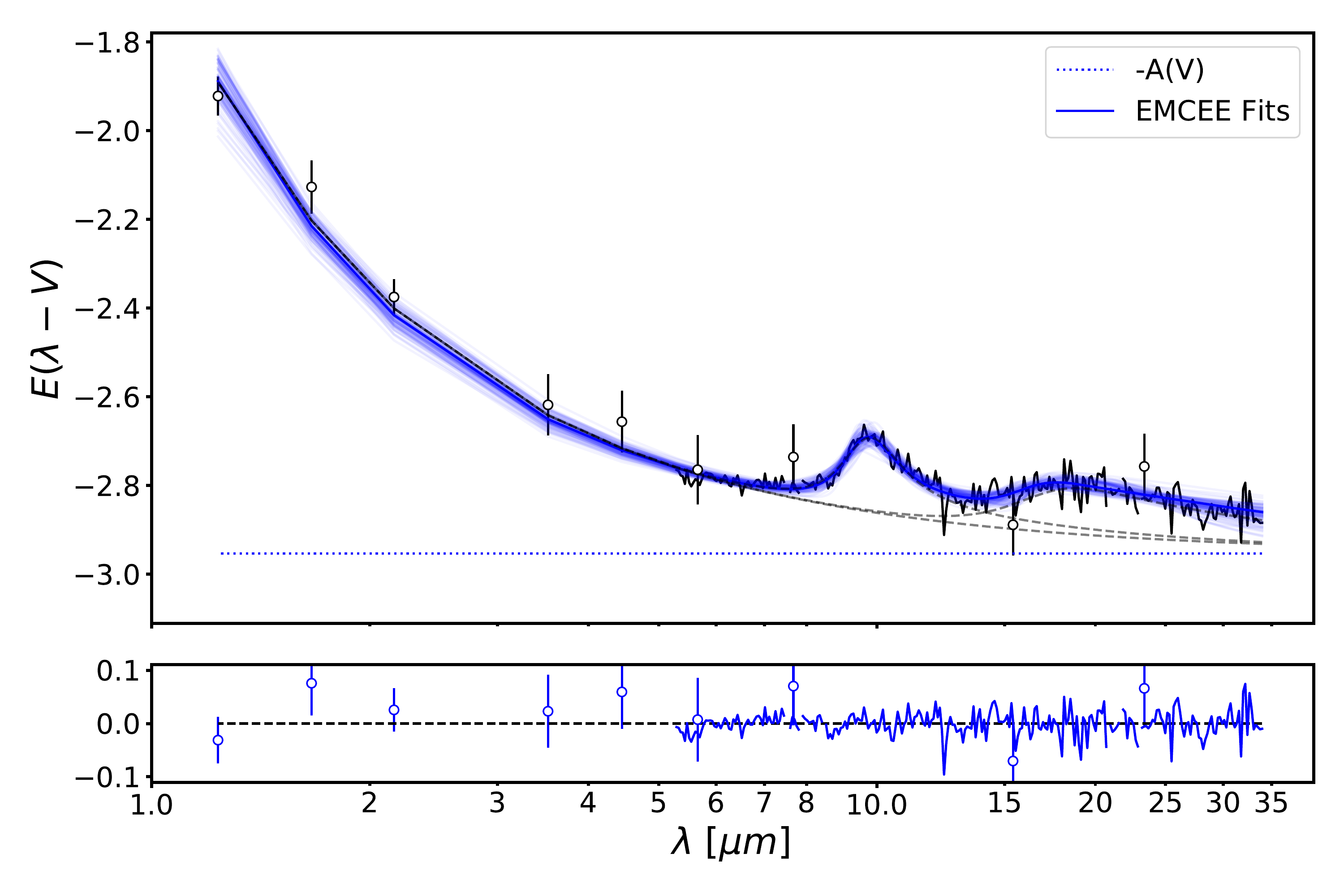}
\caption{The power law plus two modified Drudes fit to the HD112272 extinction curve is shown.
The fit uncertainties are illustrated with the cloud of blue fits.
The fitted value of $A(V)$ is shown by the horizontal dotted line.
The residuals to the fit are shown in the bottom plot.
\label{fig_fit_example} }
\end{figure}

An example of the fitting is shown in Fig.~\ref{fig_fit_example} for the sightline towards HD112272.
The power law continuum and two modified Drude profile components are shown along with the combined fit.
The fitted value of $A(V)$ is illustrated visually by the horizontal dotted line.
The residuals illustrate that the fit is reasonable.

\begin{figure*}[tbp]
\epsscale{1.2}
\plotone{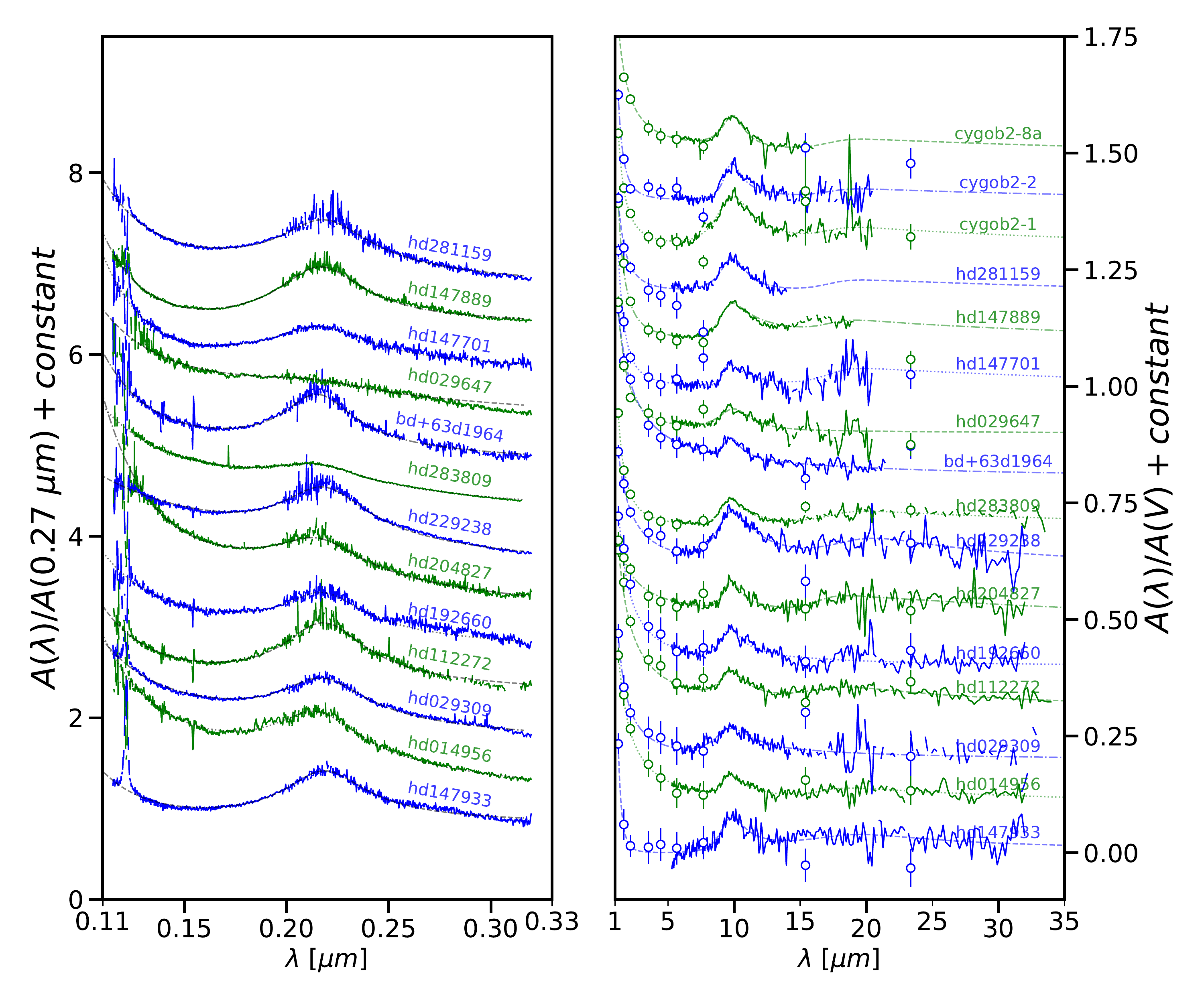}
\caption{The MIR extinction curves are shown on the right normalized to the total dust column as measured by $A(V)$.  The UV extinction curves for the same sightlines are shown on the left normalized to the total dust column, this time measured by $A(0.27~\mu m)$ to make it easy to associate the extinction curves between the two wavelength ranges.  Note the top three MIR curves do not have measured UV extinction curves.  The empirical models fit to the two wavelength regions are shown as smooth dashed lines.
\label{fig_uv_ir_alav}}
\end{figure*}

For the UV extinction curves, we use the standard FM90 empirical formulation \citep{Fitzpatrick90} to measure the 2175~\AA\ feature and strength of the far-UV rise.
The FM90 fits were done using the \alav\ extinction curves calculated using the $A(V)$ determined from the MIR fitting.
The UV fitting was done in a similar way to the MIR fitting using the `emcee' package and the `FM90' model from the `dust\_extinction' package \citep{dustextinction}.
The FM90 parameters are given in Table~\ref{tab_uv_ext_params}.

The fitted values reported in Tables~\ref{tab_mir_gen_ext_params}--\ref{tab_uv_ext_params} were computed from the marginalized posterior probability function and give the 50\% values and the asymmetric uncertainties for +34\% and -34\% from this value.
The samples used to define the fitted values were based on `emcee' fits with $2n$ walkers each with 6000 steps after a burn in of 4000 steps where $n$ is the number of fit parameters.

\begin{deluxetable}{ccccc}
\tabletypesize{\scriptsize}
\tablewidth{0pt}
\tablecaption{MIR Extinction General Parameters\label{tab_mir_gen_ext_params}}
\tablehead{\colhead{name} & \colhead{$A(V)$} & \colhead{$R(V)$} & \colhead{$B$} & \colhead{$\alpha$} \\
 & \colhead{[mag]} & \colhead{} & \colhead{[$A(1\micron)/A(V)$]} & \colhead{}}
\startdata
BD+63D1964 & $2.65^{+0.03}_{-0.03}$ & $3.40^{+0.18}_{-0.16}$ & $0.44^{+0.02}_{-0.02}$ & $0.93^{+0.08}_{-0.08}$ \\
HD014956 & $2.44^{+0.02}_{-0.02}$ & $3.05^{+0.19}_{-0.17}$ & $0.44^{+0.03}_{-0.03}$ & $1.42^{+0.15}_{-0.14}$ \\
HD029309 & $1.91^{+0.02}_{-0.02}$ & $3.83^{+0.34}_{-0.29}$ & $0.38^{+0.04}_{-0.04}$ & $1.61^{+0.33}_{-0.30}$ \\
HD029647 & $3.50^{+0.02}_{-0.02}$ & $3.44^{+0.15}_{-0.14}$ & $0.39^{+0.04}_{-0.03}$ & $1.78^{+0.27}_{-0.25}$ \\
HD112272 & $2.92^{+0.04}_{-0.03}$ & $3.33^{+0.20}_{-0.18}$ & $0.47^{+0.02}_{-0.02}$ & $1.27^{+0.15}_{-0.14}$ \\
HD147701 & $2.63^{+0.01}_{-0.01}$ & $4.02^{+0.22}_{-0.19}$ & $0.51^{+0.05}_{-0.05}$ & $2.65^{+0.38}_{-0.33}$ \\
HD147889 & $4.22^{+0.01}_{-0.01}$ & $4.18^{+0.17}_{-0.16}$ & $0.49^{+0.03}_{-0.03}$ & $2.40^{+0.21}_{-0.20}$ \\
HD147933 & $1.84^{+0.01}_{-0.01}$ & $4.32^{+0.38}_{-0.32}$ & $0.58^{+0.09}_{-0.10}$ & $4.28^{+0.51}_{-0.70}$ \\
HD192660 & $2.09^{+0.03}_{-0.03}$ & $2.40^{+0.13}_{-0.12}$ & $0.44^{+0.03}_{-0.03}$ & $1.16^{+0.16}_{-0.14}$ \\
HD204827 & $2.65^{+0.03}_{-0.02}$ & $2.48^{+0.10}_{-0.09}$ & $0.22^{+0.02}_{-0.02}$ & $1.16^{+0.24}_{-0.23}$ \\
HD229238 & $2.69^{+0.04}_{-0.03}$ & $2.75^{+0.11}_{-0.10}$ & $0.34^{+0.03}_{-0.02}$ & $1.41^{+0.26}_{-0.23}$ \\
HD281159 & $2.69^{+0.02}_{-0.01}$ & $3.14^{+0.16}_{-0.14}$ & $0.33^{+0.05}_{-0.04}$ & $2.37^{+0.54}_{-0.49}$ \\
HD283809 & $5.45^{+0.01}_{-0.01}$ & $3.41^{+0.12}_{-0.11}$ & $0.40^{+0.02}_{-0.02}$ & $2.32^{+0.20}_{-0.18}$ \\
CYGOB2-1 & $4.64^{+0.02}_{-0.02}$ & $2.87^{+0.09}_{-0.08}$ & $0.40^{+0.03}_{-0.03}$ & $2.32^{+0.24}_{-0.23}$ \\
CYGOB2-2 & $3.96^{+0.01}_{-0.01}$ & $2.93^{+0.11}_{-0.10}$ & $0.45^{+0.06}_{-0.05}$ & $3.30^{+0.42}_{-0.38}$ \\
CYGOB2-8A & $4.23^{+0.02}_{-0.02}$ & $3.16^{+0.12}_{-0.11}$ & $0.38^{+0.02}_{-0.02}$ & $1.50^{+0.14}_{-0.14}$ \\
Diffuse\tablenotemark{a} & \nodata & $3.17$ & $0.366^{+0.020}_{-0.020}$ & $1.480^{+0.046}_{-0.043}$
\enddata
\tablenotetext{a}{Obtained by fitting the diffuse average extinction extinction curve.}
\end{deluxetable}

\centerwidetable
\begin{deluxetable*}{ccccccccc}
\tabletypesize{\scriptsize}
\tablewidth{0pt}
\tablecaption{MIR Extinction Silicate Parameters\label{tab_mir_sil_ext_params}}
\tablehead{\colhead{name} & \colhead{$S_1 \times 100$} & \colhead{$\lambda_{o1}$} & \colhead{$\gamma_{o1}$} & \colhead{$a_1$} & \colhead{$S_2 \times 100$} & \colhead{$\lambda_{o2}$} & \colhead{$\gamma_{o2}$} & \colhead{$a_2$} \\
 & \colhead{[$A(\lambda_{o1})/A(V)$]} & \colhead{[$\micron$]} & \colhead{[$\micron$]} & \colhead{} & \colhead{[$A(\lambda_{o2})/A(V)$]} & \colhead{[$\micron$]} & \colhead{[$\micron$]} & \colhead{}}
\startdata
BD+63D1964 & $4.27^{+0.49}_{-0.50}$ & $9.67^{+0.20}_{-0.15}$ & $1.57^{+0.40}_{-0.30}$ & $-0.82^{+0.64}_{-0.80}$ & $0.64^{+0.54}_{-0.38}$ & $20.00$ & $13.00$ & $-0.30$ \\
HD014956 & $4.37^{+0.58}_{-0.59}$ & $9.65^{+0.19}_{-0.15}$ & $1.70^{+0.47}_{-0.35}$ & $-0.87^{+0.54}_{-0.70}$ & $2.00^{+0.71}_{-0.63}$ & $18.51^{+1.86}_{-0.97}$ & $13.00$ & $-0.67^{+0.48}_{-0.88}$ \\
HD029309 & $5.79^{+0.79}_{-0.72}$ & $9.92^{+0.29}_{-0.28}$ & $3.99^{+1.57}_{-1.00}$ & $-0.08^{+0.18}_{-0.21}$ & $1.36^{+1.11}_{-0.84}$ & $20.54^{+2.65}_{-2.91}$ & $13.00$ & $0.20^{+0.98}_{-1.09}$ \\
HD029647 & $4.66^{+0.39}_{-0.40}$ & $9.94^{+0.20}_{-0.30}$ & $2.86^{+0.84}_{-0.75}$ & $-0.04^{+0.15}_{-0.60}$ & $0.21^{+0.19}_{-0.08}$ & $20.00$ & $13.00$ & $-0.30$ \\
HD112272 & $5.39^{+0.53}_{-0.53}$ & $9.78^{+0.14}_{-0.15}$ & $2.28^{+0.53}_{-0.47}$ & $-0.32^{+0.21}_{-0.34}$ & $2.78^{+0.91}_{-0.79}$ & $18.48^{+1.23}_{-1.11}$ & $13.00$ & $-0.32^{+0.22}_{-0.61}$ \\
HD147701 & $4.71^{+0.56}_{-0.52}$ & $9.64^{+0.14}_{-0.11}$ & $1.68^{+0.36}_{-0.29}$ & $-1.25^{+0.61}_{-0.51}$ & $3.59^{+0.84}_{-0.85}$ & $20.00$ & $13.00$ & $-0.30$ \\
HD147889 & $7.57^{+0.32}_{-0.33}$ & $9.88^{+0.08}_{-0.08}$ & $2.98^{+0.29}_{-0.25}$ & $-0.49^{+0.09}_{-0.10}$ & $3.09^{+0.44}_{-0.43}$ & $20.00$ & $13.00$ & $-0.30$ \\
HD147933 & $7.89^{+0.69}_{-0.70}$ & $9.88^{+0.13}_{-0.14}$ & $2.46^{+0.55}_{-0.45}$ & $-0.18^{+0.21}_{-0.27}$ & $3.51^{+0.89}_{-0.65}$ & $16.83^{+0.87}_{-0.57}$ & $13.00$ & $-0.45^{+0.25}_{-0.47}$ \\
HD192660 & $6.55^{+0.67}_{-0.66}$ & $9.68^{+0.25}_{-0.20}$ & $2.59^{+0.63}_{-0.54}$ & $-0.67^{+0.40}_{-0.62}$ & $2.18^{+0.97}_{-1.00}$ & $19.28^{+1.30}_{-0.75}$ & $13.00$ & $-1.13^{+0.61}_{-0.59}$ \\
HD204827 & $5.75^{+0.61}_{-0.59}$ & $9.74^{+0.14}_{-0.14}$ & $1.52^{+0.33}_{-0.27}$ & $-0.60^{+0.38}_{-0.68}$ & $3.11^{+0.82}_{-0.68}$ & $18.24^{+1.55}_{-0.78}$ & $13.00$ & $-0.76^{+0.51}_{-0.78}$ \\
HD229238 & $10.50^{+0.65}_{-0.64}$ & $9.83^{+0.13}_{-0.14}$ & $3.24^{+0.48}_{-0.42}$ & $-0.29^{+0.10}_{-0.14}$ & $3.60^{+1.37}_{-0.81}$ & $19.39^{+2.11}_{-1.21}$ & $13.00$ & $-0.55^{+0.42}_{-0.86}$ \\
HD281159 & $7.27^{+0.56}_{-0.51}$ & $9.87^{+0.10}_{-0.11}$ & $2.33^{+0.35}_{-0.31}$ & $-0.06^{+0.12}_{-0.16}$ & $2.50$ & $20.00$ & $13.00$ & $-0.30$ \\
HD283809 & $5.71^{+0.26}_{-0.25}$ & $9.73^{+0.07}_{-0.08}$ & $1.97^{+0.18}_{-0.17}$ & $-0.40^{+0.13}_{-0.17}$ & $2.73^{+0.31}_{-0.28}$ & $19.85^{+1.08}_{-1.04}$ & $13.00$ & $-0.35^{+0.16}_{-0.32}$ \\
CYGOB2-1 & $10.54^{+0.35}_{-0.36}$ & $9.90^{+0.06}_{-0.06}$ & $3.42^{+0.29}_{-0.26}$ & $-0.13^{+0.04}_{-0.04}$ & $2.86^{+0.64}_{-0.62}$ & $20.00$ & $13.00$ & $-0.30$ \\
CYGOB2-2 & $7.82^{+0.47}_{-0.48}$ & $9.82^{+0.10}_{-0.11}$ & $1.68^{+0.19}_{-0.17}$ & $-0.47^{+0.19}_{-0.27}$ & $1.94^{+0.56}_{-0.52}$ & $20.00$ & $13.00$ & $-0.30$ \\
CYGOB2-8A & $6.86^{+0.32}_{-0.30}$ & $9.91^{+0.07}_{-0.06}$ & $2.02^{+0.18}_{-0.17}$ & $0.07^{+0.10}_{-0.10}$ & $2.50$ & $20.00$ & $13.00$ & $-0.30$ \\
Diffuse\tablenotemark{a} & $6.893^{+0.152}_{-0.155}$ & $9.865^{+0.037}_{-0.037}$ & $2.507^{+0.123}_{-0.114}$ & $-0.232^{+0.058}_{-0.059}$ & $2.684^{+0.171}_{-0.138}$ & $19.973^{+0.709}_{-0.607}$ & $16.989^{+1.945}_{-1.980}$ & $-0.273^{+0.108}_{-0.130}$ \\
\enddata
\tablenotetext{a}{Obtained by fitting the diffuse average extinction extinction curve.}
\end{deluxetable*}

\centerwidetable
\begin{deluxetable*}{ccccccc}
\tabletypesize{\scriptsize}
\tablewidth{0pt}
\tablecaption{UV Extinction Parameters\label{tab_uv_ext_params}}
\tablehead{\colhead{name} & \colhead{$C1$} & \colhead{$C2$} & \colhead{$C3$} & \colhead{$C4$} & \colhead{$x_o$} & \colhead{$\gamma$} \\
 & \colhead{[$A(\lambda)/A(V)$]} & \colhead{[$A(\lambda)/A(V)$]} & \colhead{[$A(\lambda)/A(V)$]} & \colhead{[$A(\lambda)/A(V)$]} & \colhead{[$\micron^{-1}$]} & \colhead{[$\micron^{-1}$]}}
\startdata
BD+63D1964 & $1.180^{+0.010}_{-0.010}$ & $0.173^{+0.002}_{-0.002}$ & $0.934^{+0.011}_{-0.011}$ & $0.173^{+0.004}_{-0.004}$ & $4.640^{+0.001}_{-0.001}$ & $0.914^{+0.005}_{-0.005}$ \\
HD014956 & $0.556^{+0.009}_{-0.009}$ & $0.370^{+0.002}_{-0.002}$ & $1.519^{+0.017}_{-0.016}$ & $0.184^{+0.006}_{-0.006}$ & $4.645^{+0.001}_{-0.001}$ & $1.143^{+0.005}_{-0.005}$ \\
HD029309 & $0.903^{+0.008}_{-0.007}$ & $0.224^{+0.001}_{-0.001}$ & $1.005^{+0.011}_{-0.011}$ & $0.093^{+0.004}_{-0.004}$ & $4.584^{+0.001}_{-0.001}$ & $1.067^{+0.005}_{-0.005}$ \\
HD029647 & $1.142^{+0.011}_{-0.012}$ & $0.259^{+0.002}_{-0.002}$ & $0.552^{+0.008}_{-0.008}$ & $0.100^{+0.013}_{-0.013}$ & $4.654^{+0.006}_{-0.006}$ & $1.499^{+0.001}_{-0.001}$ \\
HD112272 & $1.236^{+0.014}_{-0.013}$ & $0.146^{+0.002}_{-0.002}$ & $1.371^{+0.020}_{-0.019}$ & $0.135^{+0.004}_{-0.004}$ & $4.575^{+0.001}_{-0.001}$ & $1.033^{+0.007}_{-0.007}$ \\
HD147701 & $1.203^{+0.013}_{-0.013}$ & $0.103^{+0.002}_{-0.002}$ & $0.981^{+0.021}_{-0.020}$ & $0.217^{+0.004}_{-0.004}$ & $4.616^{+0.001}_{-0.002}$ & $1.230^{+0.009}_{-0.009}$ \\
HD147889 & $1.293^{+0.004}_{-0.004}$ & $0.040^{+0.001}_{-0.001}$ & $1.167^{+0.008}_{-0.008}$ & $0.194^{+0.002}_{-0.002}$ & $4.612^{+0.000}_{-0.000}$ & $1.088^{+0.003}_{-0.003}$ \\
HD147933 & $1.271^{+0.009}_{-0.009}$ & $0.053^{+0.001}_{-0.002}$ & $0.917^{+0.011}_{-0.011}$ & $0.084^{+0.003}_{-0.003}$ & $4.567^{+0.001}_{-0.001}$ & $1.025^{+0.005}_{-0.005}$ \\
HD192660 & $1.135^{+0.017}_{-0.016}$ & $0.265^{+0.003}_{-0.003}$ & $1.008^{+0.021}_{-0.020}$ & $0.122^{+0.009}_{-0.009}$ & $4.587^{+0.001}_{-0.001}$ & $0.996^{+0.008}_{-0.008}$ \\
HD204827 & $0.582^{+0.011}_{-0.011}$ & $0.458^{+0.002}_{-0.002}$ & $1.245^{+0.021}_{-0.022}$ & $0.392^{+0.012}_{-0.012}$ & $4.631^{+0.001}_{-0.001}$ & $1.123^{+0.009}_{-0.009}$ \\
HD229238 & $0.786^{+0.007}_{-0.007}$ & $0.331^{+0.001}_{-0.001}$ & $1.582^{+0.014}_{-0.014}$ & $0.009^{+0.005}_{-0.005}$ & $4.572^{+0.001}_{-0.001}$ & $1.098^{+0.005}_{-0.005}$ \\
HD281159 & $1.091^{+0.013}_{-0.013}$ & $0.199^{+0.002}_{-0.002}$ & $1.451^{+0.024}_{-0.025}$ & $0.159^{+0.004}_{-0.004}$ & $4.567^{+0.001}_{-0.001}$ & $1.133^{+0.010}_{-0.010}$ \\
HD283809 & $0.890^{+0.003}_{-0.003}$ & $0.241^{+0.001}_{-0.001}$ & $0.574^{+0.007}_{-0.007}$ & $0.073^{+0.004}_{-0.004}$ & $4.635^{+0.001}_{-0.001}$ & $1.172^{+0.005}_{-0.005}$ \\
Diffuse\tablenotemark{a} & $1.0238^{+0.0212}_{-0.0213}$ & $0.2128^{+0.0058}_{-0.0058}$ & $1.2272^{+0.0707}_{-0.0648}$ & $0.1656^{+0.0201}_{-0.0209}$ & $4.5979^{+0.0053}_{-0.0054}$ & $1.0755^{+0.0283}_{-0.0274}$
\enddata
\tablenotetext{a}{Obtained by fitting the diffuse average extinction extinction curve.}
\end{deluxetable*}

The UV and MIR extinction curves and fits for all our sightlines are shown in Fig.~\ref{fig_uv_ir_alav}.
We have used the fitted values of $A(V)$ for each sightline to transform the curves from unnormalized \elv\ curves to dust column normalized \alav\ curves.
Overall, the MIR fits are good, especially when the main goal is extrapolating the curves to infinite wavelength to derive $A(V)$.
The MIR fits do clearly illustrate that there is significant instrumental noise present in the longer wavelength portions of the curves.
This is not unexpected given the sensitivity of the IRS instrument and flux from stars is dropping rapidly towards longer wavelengths.

\section{Results}
\label{sec_results}

\begin{figure}[tbp]
\epsscale{1.2}
\plotone{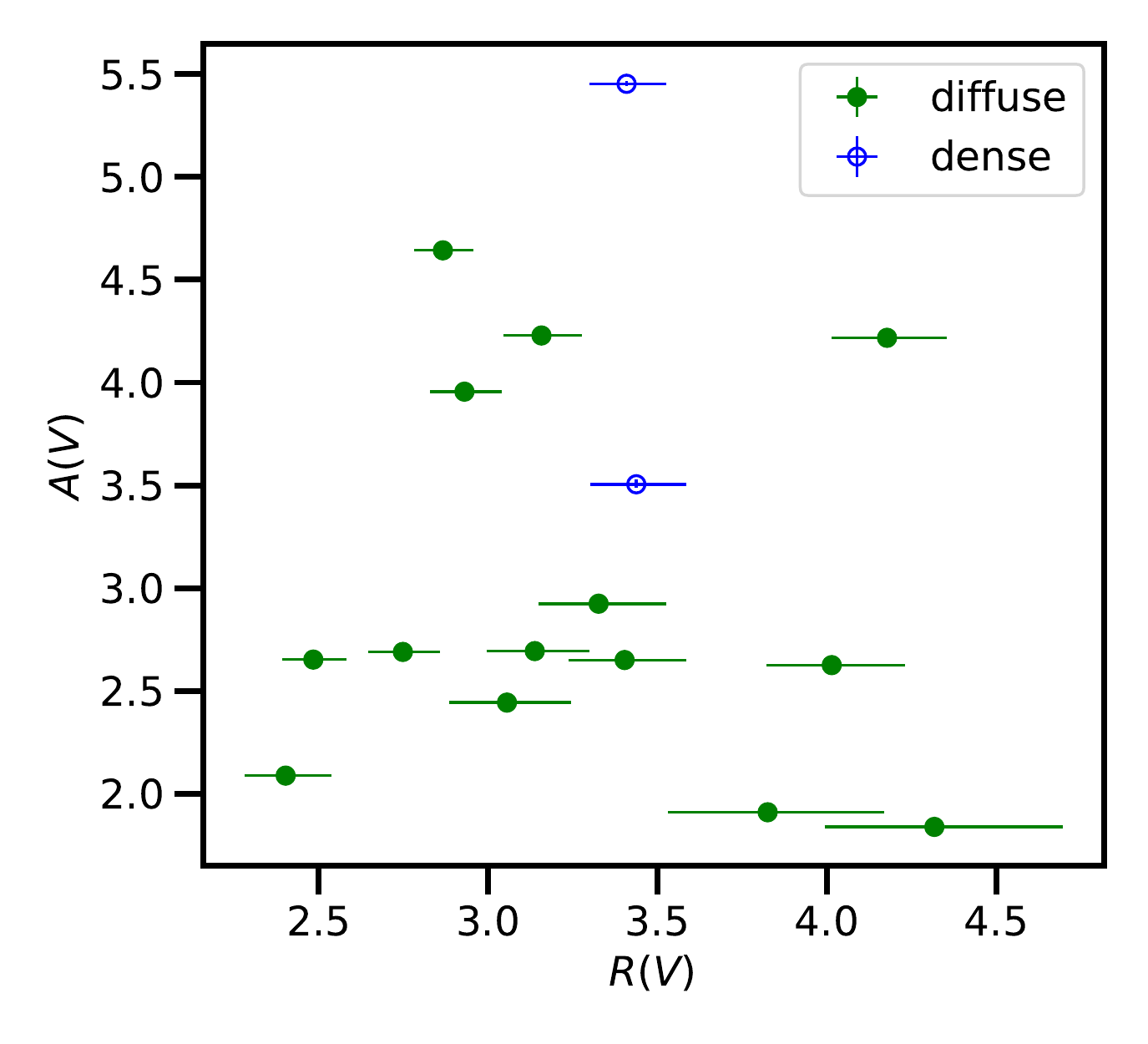}
\caption{The $A(V)$ versus $R(V)$ plot illustrates the general properties of our sample showing that it probes a range of dust grain sizes (as probed by $R(V)$) and dust columns ($A(V)$).
\label{fig_sampprop}}
\end{figure}

The general properties of our sample are illustrated in Fig.~\ref{fig_sampprop} showing the distribution of $A(V)$ versus $R(V)$.
The $A(V)$ values measure the dust column and range from 1.8--5.5~mag.
This sample has $R(V)$ values from 2.4--4.3 and thus encompasses a large fraction of the Milky Way observed range of 2--6 \citep{Valencic04, Fitzpatrick07}.

The diffuse or dense nature of each sightline is given in Table~\ref{tab_red_sample}.
The diffuse versus dense nature is based on the absence or presence of ice absorption at 3.0~\micron, respectively.
Only sightlines with $A(V) > 3$ have enough shielding to allow for ice formation and show the ice absorption feature \citep{Boogert15}.
Hence sightlines with $A(V) < 3$ are considered diffuse even without spectroscopic measurements in the 3~\micron\ region.
Only the HD29647 and HD283809 sightlines have detected 3.0~\micron\ ice features \citep{Whittet88, Smith93} indicating they are definite dense sightlines.
For the other $A(V) > 3$ sightlines, we designate them as diffuse for the following reasons.
For HD147889, ice was not detected along this sightline \citep{Tanaka90}.
The Cyg~OB2 1, 2, \& 8A sightlines are in the Cyg~OB2 cluster and ice absorption is not detected in a well-studied member of the same cluster with higher extinction \citep[Cyg OB2 12][]{Gillett75, Whittet15}.

In general, most of our sample reveal similar overall extinction curves as shown in Fig.~\ref{fig_uv_ir_alav}.
In the UV, the HD29647 and HD283809 sightlines show significantly weaker 2175~\AA\ bump features than are seen in the rest of the sample.
These variations are attributed to dense versus diffuse material along the HD29647 and HD283809 sightlines \citep{Whittet04}.
In the NIR/MIR, the BD+63D1964 sightline shows a steeper slope and the HD283809 sightline a much shallower slope when compared to the rest of the sample.
Clear variations in the strength of the 10~\micron\ silicate feature are seen and investigated in more detail in Sec.~\ref{sec_sil_features}.
The NIR/MIR continuum variations may be due to probing diffuse versus dense material, but they may also result from challenges in measuring extinctions in this wavelength range.

\subsection{Average Diffuse Extinction}

\begin{figure*}[tbp]
\epsscale{1.15}
\plotone{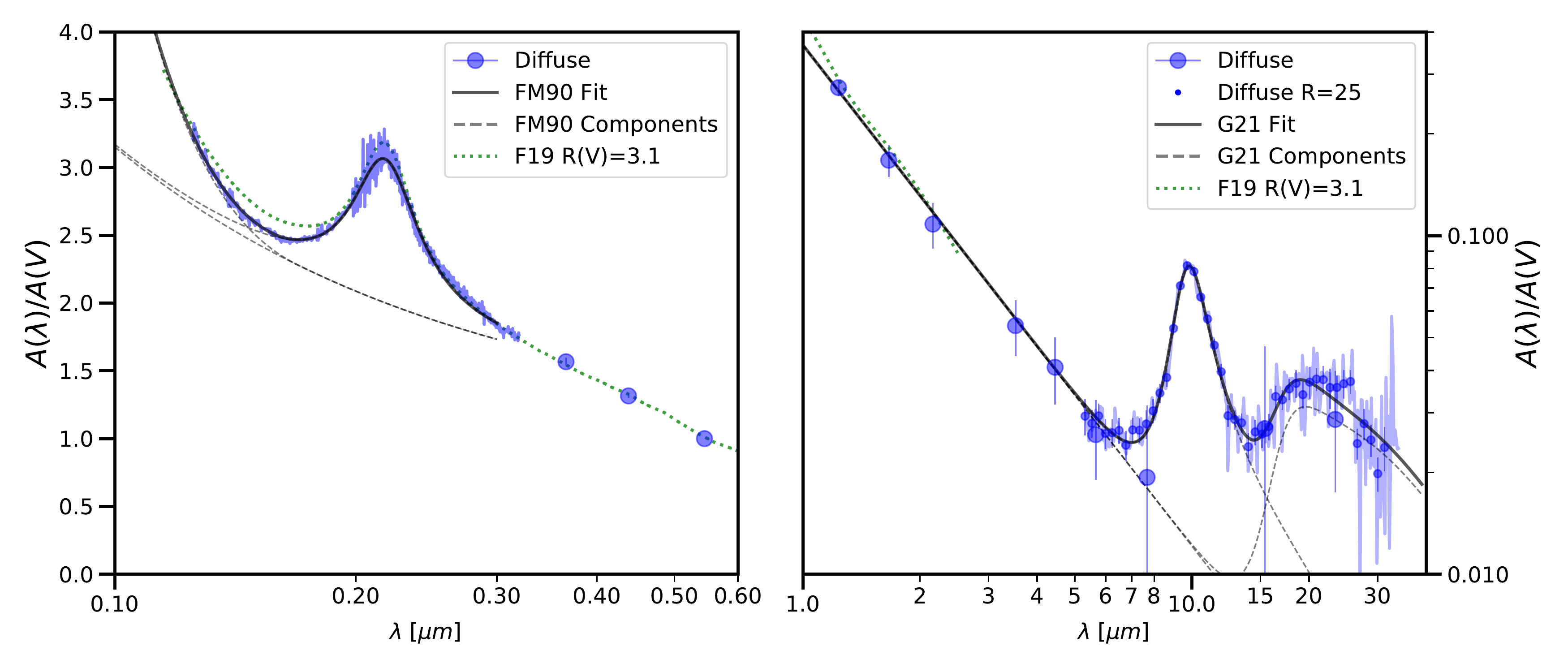}
\caption{The average extinction of all of the diffuse sightlines in our sample is plotted along with the best fits and components of the best fits.
The IRS portion of the curve is shown at full resolution and rebinned at a resolution of $R=25$.
The average Milky Way diffuse UV--NIR curve is very similar to the F19 $R(V) = 3.1$ curve \citep{Fitzpatrick19}.
The UV continuum fitted component is slightly curved when plotted versus $\lambda$, but is linear versus $\lambda^{-1}$ \citep{Fitzpatrick90}.
\label{fig_diffuse_aveext}}
\end{figure*}

\begin{deluxetable}{cccc|cccc}
\tabletypesize{\scriptsize}
\tablewidth{0pt}
\tablecaption{Average Diffuse Milky Way Extinction \label{tab_ave_mwext}}
\tablehead{\colhead{$\lambda$} & \colhead{$k(\lambda)$} & \colhead{unc} & \colhead{G21 fit} & \colhead{$\lambda$} & \colhead{$k(\lambda)$} & \colhead{unc} & \colhead{G21 fit} \\
\colhead{[$\micron$]} & \multicolumn{3}{c}{[$A(\lambda)$/$A(V)$]} & \colhead{[$\micron$]} & \multicolumn{3}{c}{[$A(\lambda)/A(V)$]} }
\startdata
\multicolumn{4}{c}{Photometric}   & 10.54 & 0.0659 & 0.0022 & 0.0673 \\
1.24 & 0.2739 & 0.0149 & 0.2678  & 10.97 & 0.0568 & 0.0019 & 0.0553 \\
1.66 & 0.1671 & 0.0181 & 0.1726  & 11.42 & 0.0475 & 0.0022 & 0.0455 \\
2.16 & 0.1082 & 0.0167 & 0.1172  & 11.89 & 0.0396 & 0.0022 & 0.0382 \\
3.52 & 0.0542 & 0.0102 & 0.0569  & 12.38 & 0.0294 & 0.0022 & 0.0330 \\
4.45 & 0.0409 & 0.0092 & 0.0404  & 12.89 & 0.0286 & 0.0019 & 0.0292 \\
5.66 & 0.0258 & 0.0068 & 0.0291  & 13.42 & 0.0280 & 0.0018 & 0.0267 \\
7.67 & 0.0193 & 0.0122 & 0.0264  & 13.97 & 0.0238 & 0.0023 & 0.0253 \\
15.40 & 0.0269 & 0.0202 & 0.0264 & 14.55 & 0.0263 & 0.0023 & 0.0249 \\
23.36 & 0.0287 & 0.0112 & 0.0327 & 15.14 & 0.0259 & 0.0023 & 0.0257 \\
\multicolumn{4}{c}{Spectroscopic} & 15.77 & 0.0273 & 0.0024 & 0.0278 \\
5.31 & 0.0293 & 0.0036 & 0.0316  & 16.41 & 0.0335 & 0.0026 & 0.0307 \\
5.53 & 0.0279 & 0.0027 & 0.0299  & 17.09 & 0.0328 & 0.0022 & 0.0337 \\
5.76 & 0.0294 & 0.0026 & 0.0285  & 17.79 & 0.0353 & 0.0025 & 0.0361 \\
6.00 & 0.0260 & 0.0026 & 0.0272  & 18.52 & 0.0366 & 0.0035 & 0.0373 \\
6.24 & 0.0262 & 0.0025 & 0.0261  & 19.28 & 0.0339 & 0.0031 & 0.0375 \\
6.50 & 0.0266 & 0.0024 & 0.0252  & 20.08 & 0.0369 & 0.0040 & 0.0369 \\
6.77 & 0.0240 & 0.0025 & 0.0247  & 20.90 & 0.0377 & 0.0029 & 0.0359 \\
7.04 & 0.0267 & 0.0023 & 0.0245  & 21.76 & 0.0376 & 0.0037 & 0.0348 \\
7.33 & 0.0266 & 0.0022 & 0.0249  & 22.66 & 0.0356 & 0.0048 & 0.0336 \\
7.63 & 0.0278 & 0.0027 & 0.0262  & 23.59 & 0.0356 & 0.0030 & 0.0324 \\
7.95 & 0.0304 & 0.0025 & 0.0289  & 24.56 & 0.0365 & 0.0035 & 0.0313 \\
8.27 & 0.0343 & 0.0023 & 0.0338  & 25.57 & 0.0371 & 0.0031 & 0.0301 \\
8.62 & 0.0381 & 0.0024 & 0.0421  & 26.62 & 0.0243 & 0.0026 & 0.0290 \\
8.97 & 0.0532 & 0.0021 & 0.0548  & 27.71 & 0.0278 & 0.0030 & 0.0279 \\
9.34 & 0.0712 & 0.0022 & 0.0704  & 28.85 & 0.0249 & 0.0027 & 0.0268 \\
9.72 & 0.0816 & 0.0023 & 0.0807  & 30.04 & 0.0198 & 0.0023 & 0.0257 \\
10.12 & 0.0784 & 0.0024 & 0.0784 & 31.27 & 0.0237 & 0.0035 & 0.0246
\enddata
\end{deluxetable}

The average extinction for the diffuse ISM sightlines in our sample is shown in Fig.~\ref{fig_diffuse_aveext} along with the MIR and UV fits.
The average was computed without any weighting as the propagated uncertainties account only for the measurement and calibration uncertainty and do not include systematic uncertainties due to the mismatches between the reddened and comparison stars.
Given that the mismatch systematic effects should not be correlated between extinction curves, they average out.
The uncertainties on the average were computed as the standard deviation of the mean and were seen visually to match the spectral variations in the average (e.g., Sec.~\ref{sec_faint_features}).
This average probes a summed dust column with $A(V) \sim 42$~mag, much higher than any of the individual curves enabling measurements of weaker structures and features.
The MIR and UV fit parameters for the average are included in Tables~\ref{tab_mir_gen_ext_params}--\ref{tab_uv_ext_params}.
The Milky Way diffuse UV--NIR extinction curve derived from a larger sample of diffuse sightlines \citep{Fitzpatrick19} is included in Fig.~\ref{fig_diffuse_aveext} and is very similar to our average.
This supports the nature of our diffuse sightlines and that our average is a reasonable measure of the diffuse dust in the Milky Way.
We searched for the signature of $R(V)$ dependent variations in our sample MIR extinction curves and found no significant variations.

The average NIR--MIR diffuse Milky Way extinction values are given in Table~\ref{tab_ave_mwext} for the photometric bands and for a resolution $R = 25$ for the IRS portion.
We also provide tabulated values for the G21 fit.
Note that the average extinction curve below 5~\micron\ is derived using photometric measurements and, hence, can be used directly for sources with similar stellar spectral shapes.
For sources with different spectral shapes, users should account for the different weighting across the photometric filters both in the extinction curve and in their source \citep[see discussion in][]{Fitzpatrick99review}.
The average extinction curve above 5~\micron\ is derived from spectroscopic data and such effects are negligible allowing the average extinction curve to be used for all observations without any such corrections.
The full resolution curve, rebinned version, and fit are available as part of the `G21\_MWAvg' model in the `dust\_extinction' package \citep{dustextinction}.

\subsubsection{Comparison to Previous Measurements}

\begin{figure*}[tbp]
\epsscale{1.15}
\plotone{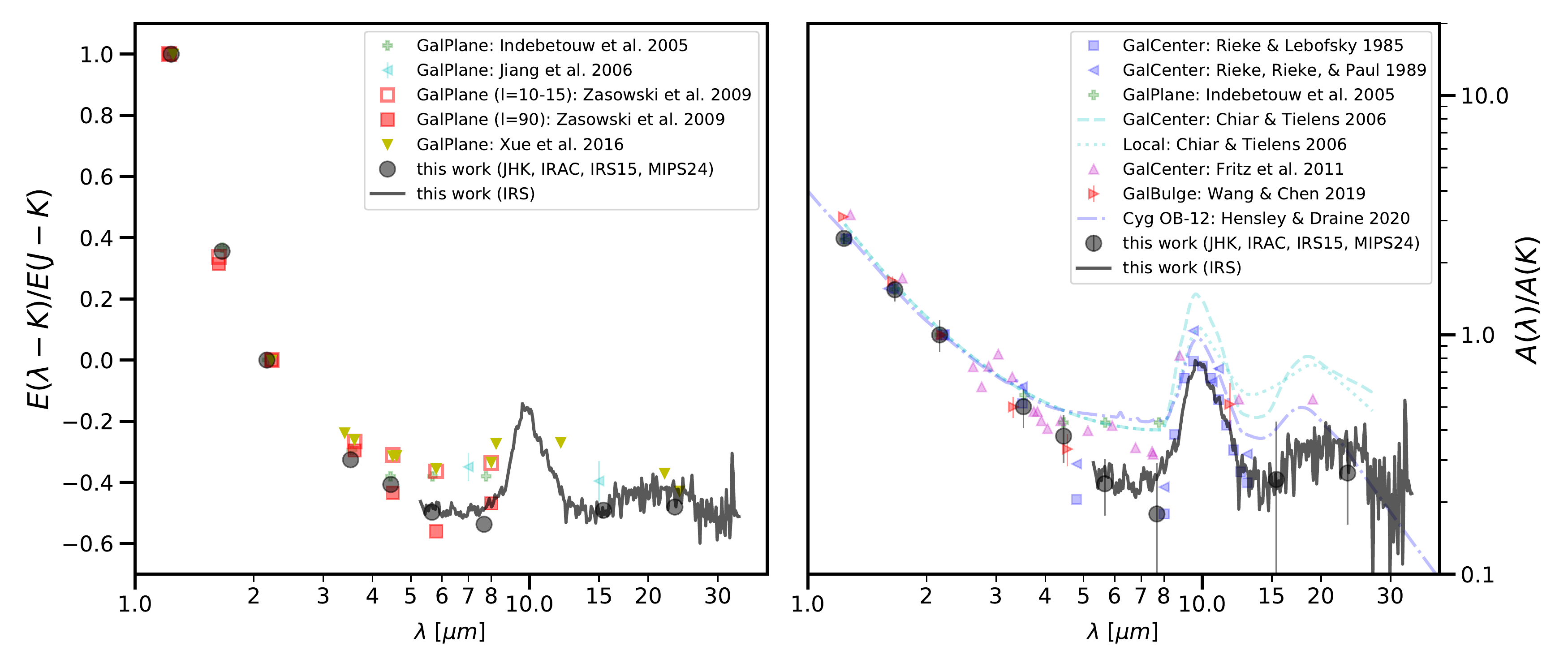}
\caption{Our results are compared to a representative set from the literature both as relative \citep[left, \elkejk;][]{Indebetouw05, Jiang06, Zasowski09, Xue16} and absolute \citep[right, \alak;][]{Rieke85, Rieke89, Indebetouw05, Chiar06, Fritz11, Wang19, Hensley20} measurements.
Real variations are seen in the relative measurements indicating different MIR extinctions between the diffuse and dense ISM.
The variations in the absolute measurements are likely due to a combination of real variations and uncertainties in deriving the total column normalizations.
Note that the \citet{Lutz99} work presented preliminary results that were finalized in \citet{Fritz11}.
\label{fig_litcom}}
\end{figure*}

Previous measurements of the average MIR extinction have generally used sightlines with large dust columns to amplify the relatively weak MIR extinction signal, generally limiting the measurements to extinctions $\geq$1~\micron.
As such, Fig.~\ref{fig_litcom} compares the average MIR extinction determined in this work with literature results for both relative, \elkejk, and absolute, \alak, extinction measurements normalized at NIR wavelengths.
Overall, our measurements are consistent with \citet{Rieke85}, \citet{Rieke89}, $l = 90$ \citet{Zasowski09}, and \citet{Wang19} measurements and lower than the rest of the literature measurements.

In the case of the \elkejk\ measurements, these differences are likely due to the relative contribution between diffuse and dense material in the sightlines studied.
This can be seen from the \citet{Zasowski09} measurements that encompass the full range of values with higher extinctions seen towards the Galactic Center ($l = 10-15$) and lower extinctions well away from the Galactic Center ($l = 90$).
\cite{Zasowski09} interpret this as less dense material in the $l = 90$ case compared to the $l = 10-15$ case and this would be consistent with much of the previous work probing more dense material given their biases towards high dust column sightlines.

In the case of the \alak\ measurements, some of the differences can be due to uncertainties in deriving the total column.
The absolute level for the \citet{Chiar06} curves was set by fiat to that of \citet{Indebetouw05} and \citet{Lutz99} and, hence, their overall high MIR extinction may just be due to this normalization.
Similarly, \citet{Hensley20} used \citet{Fitzpatrick19} to set the absolute level of the Cyg~OB2~12 extinction.

However, the differences between the \citet{Rieke89} and \citet{Fritz11} results are enigmatic.
The Br$\alpha$/Br$\gamma$ ratio for the former reference agrees well with the measurements by \citet{1987ApJ...320..570W}; between these references the ratio is measured for eight sightlines (one twice) and the (weighted) average Br$\alpha$/Br$\gamma$ ratio is 16 $\pm$ 1.
Some of the flux in the beams used may originate from luminous early-type stars, which typically have a Br$\gamma$ EW of order 6 $\times$ 10$^{-4}$ $\mu$m \citep{1996ApJS..107..281H}.
The result could be that the ratio is overestimated by about 5\%.
\citet{Fritz11} determined a ratio value of 11.6 $\pm$ 2 from the ISO spectrum (for Br$\alpha$) and a ground-based spectral image (for Br$\gamma$).
The difference in these two measurement approaches suggests that the best value is toward the large end of the ones permitted by \citet{Fritz11}, and that the Galactic Center extinction in Figure~\ref{fig_litcom} probably lies between that shown for \citet{Fritz11} and ours (and the point from \citet{Rieke89}.

Only \citet{Rieke85}, \citet{Rieke89}, \citet{Chiar06}, and \citet{Hensley20} have enough measurements in the 10~\micron\ region to investigate variations in the 10~\micron\ silicate feature.
Our average silicate feature profile closely matches that of \citet{Rieke85}.
The central strength is similar to the \citet{Rieke89}, \citet{Chiar06} Local, and \citep{Hensley20} Cyg~OB2~12 curves.
The \citet{Chiar06} Galactic Center silicate profiles have a larger amplitude than either our or the \citet{Rieke85} and \citet{Rieke89} profiles, likely due to the assumptions used by \citet{Chiar06} in constructing their Galactic Center curve as compared to the more direct measurements of \citet{Rieke85} and \citet{Rieke89}.

\subsubsection{Comparison to Dust Grain Models}

\begin{figure}[tbp]
\epsscale{1.15}
\plotone{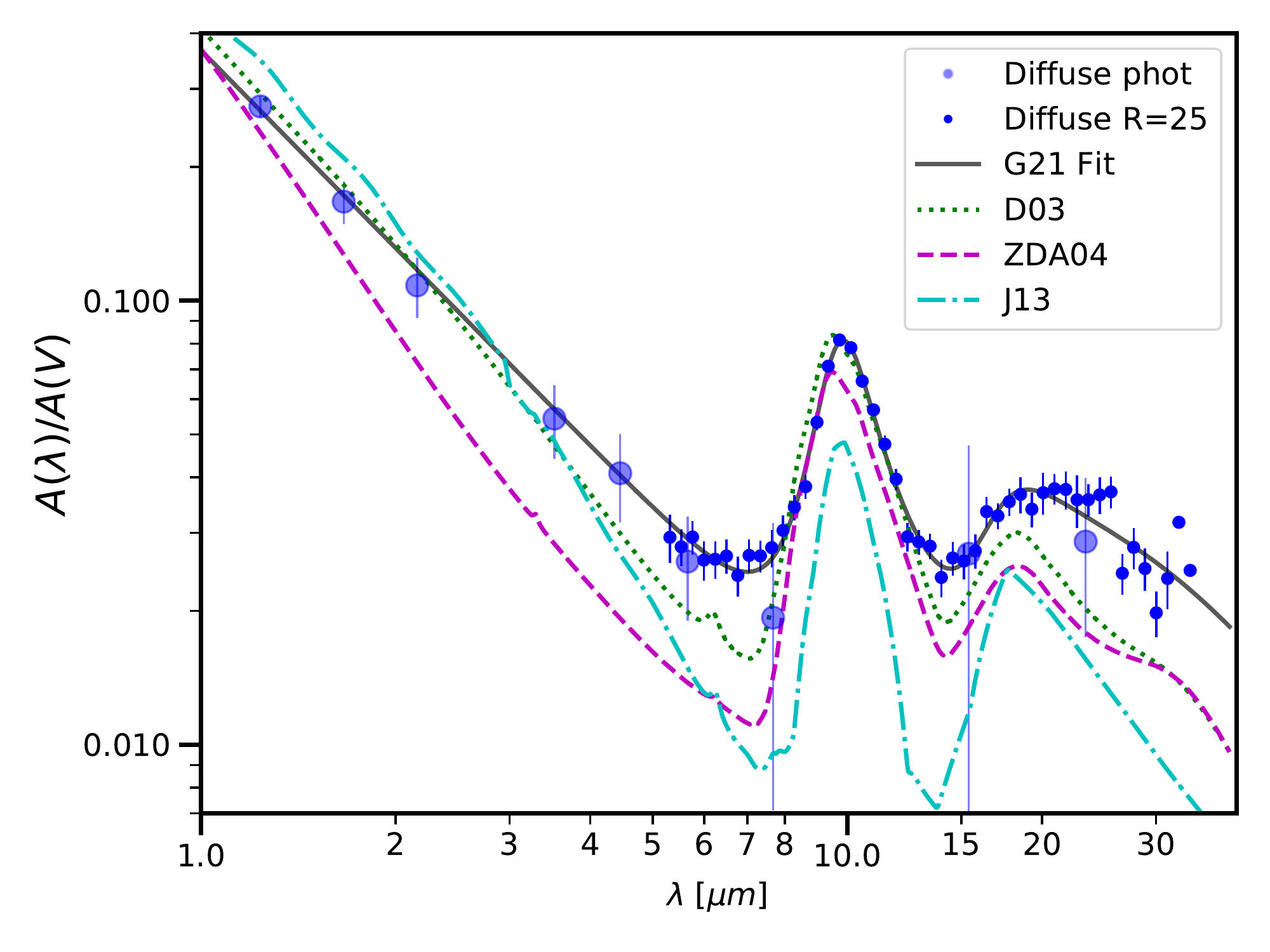}
\caption{The diffuse average is compared to the D03 \citep{Draine03review, Draine03b}, ZDA04 \citep{Zubko04}, and J13 \citep{Jones13} dust grain models for the diffuse ISM with $R(V) = 3.1$.
The WD01 \citep{Weingartner01} is very similar to the D03 model and is not shown.
All the models predict extinctions lower than our new measured average.
\label{fig_modcom}}
\end{figure}

In Fig.~\ref{fig_modcom}, our diffuse average extinction curve is compared to the D03 \citep{Draine03review, Draine03b}, ZDA04 \citep{Zubko04}, and J13 \citep{Jones13} diffuse ISM dust grain models.
Overall, the dust grain models are significantly below the diffuse curve except in the center of the 10~\micron\ silicate feature for the D03 model.
For the D03 model, this is not surprising as it was fit to the \citet{Rieke85} curve that has overall a bit lower extinction than the diffuse curve except for in the 10~\micron\ silicate feature (see Fig.~\ref{fig_litcom}).
The other dust grain models were not constrained to fit the MIR extinction and, therefore, agreement is not expected.

\subsection{Silicate Features}
\label{sec_sil_features}

\begin{figure*}[tbp]
\epsscale{1.2}
\plotone{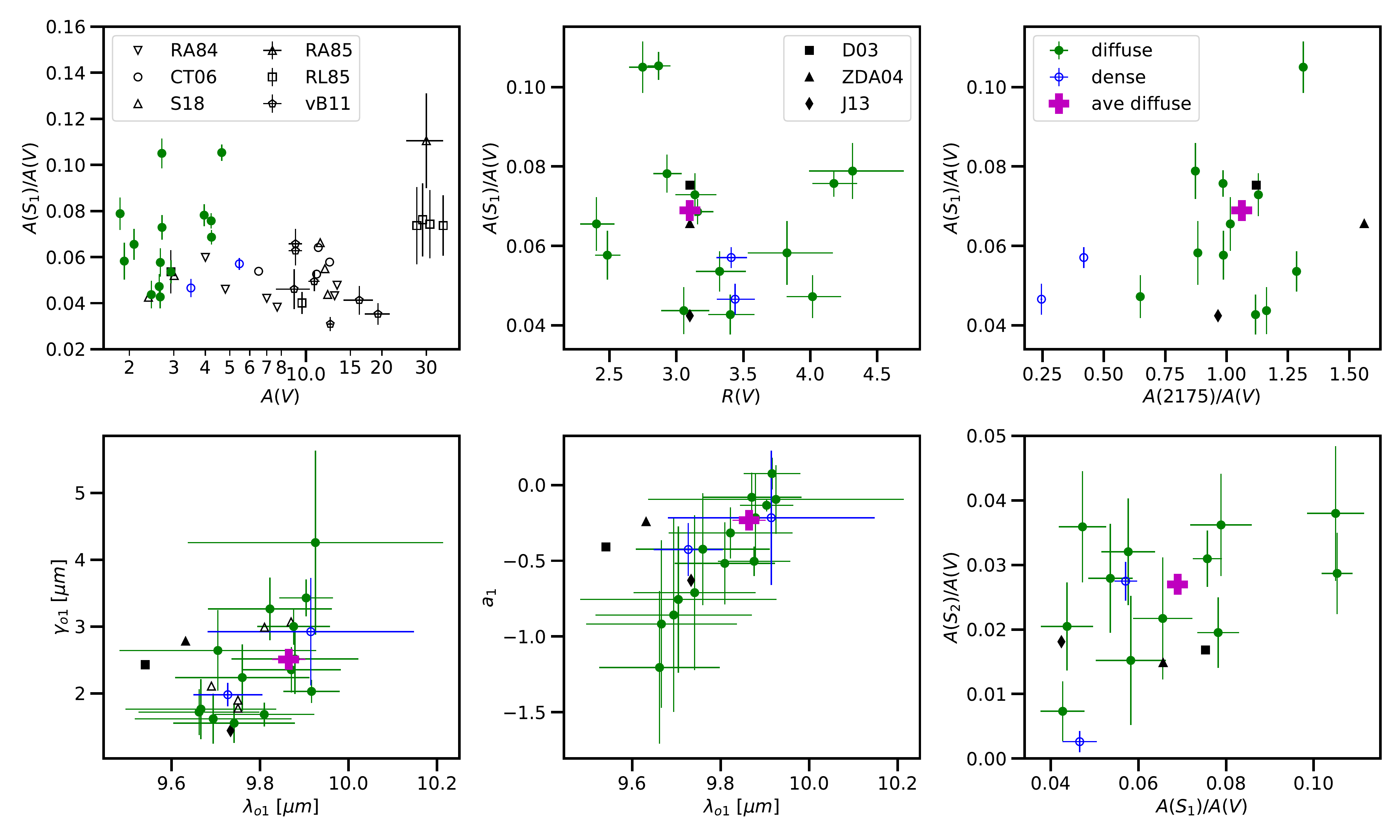}
\caption{The strength of the 10~\micron\ silicate feature ($A(S1)/A(V)$) is plotted versus the total dust column ($A(V)$), a measure of the average grain size ($R(V)$), and the strength of the 2175~\AA\ bump $A(2175)/A(V)$ along the top row of plots.
The bottom row of plots gives the most significant correlations between the 10 and 20~\micron\ feature properties.
Previous observations given in the upper left plot are RA84 \citep{Roche84}, RA85 \citep{Roche85}, RL85 \citep{Rieke85}, CT06 \citep{Chiar06}, vB11 \citep{vanBreemen11}, and S18 \citep{Shao18}.
The dust grain model points given in other plots are D03 \citep{Draine03review, Draine03b}, ZDA04 \citep{Zubko04}, and J13 \citep{Jones13} and were determined by performing the same fitting as the observational data.
\label{fig_silcor}}
\end{figure*}

The detailed properties of the silicate features and other properties of the sightlines were investigated using the parameters given in Tables~\ref{tab_mir_gen_ext_params}, \ref{tab_mir_sil_ext_params}, and \ref{tab_uv_ext_params}.
The plots in the upper row of Fig.~\ref{fig_silcor} show that the strength of the 10~\micron\ silicate feature does not correlate with $A(V)$, $R(V)$, or the 2175~\AA\ extinction bump strength.
The 10~\micron\ feature strength is characterized by the central amplitude in these plots and we found that using the integrated area resulted in similar plots.
$A(V)$ is a measure of the dust column and over a factor of two variation in the strength is seen for similar values of $A(V)$.
This variation is similar to what has been seen in previous works \citep{Roche84, Roche85, Rieke85, Chiar06, vanBreemen11, Shao18} with the main difference that our sample has larger variations at lower $A(V)$ values.
We do not see any correlation between a measure of the average grain size $R(V)$ and the 10~\micron\ silicate feature strength.
Finally, the 10~\micron\ silicate feature is not correlated with the 2175~\AA\ bump feature.
This is the first time that the correlation between these two features has been directly tested observationally.
Such a lack of correlation indicates that these two features are not due to the same type of dust grains and this agrees with the identification of the 10~\micron\ feature with silicate grains and the 2175~\AA\ feature with carbonaceous grains.
Lack of correlation could additionally indicate that environmental modification and/or formation processes modify the responsible grain populations independently.

The plots in the bottom row of Fig.~\ref{fig_silcor} show the three best correlations among and between the silicate feature properties.
We find that the 10~\micron\ central wavelength $\lambda_{o1}$, width $\gamma_{o1}$, and asymmetry $a_1$ are well correlated with each other.
This indicates that the 10~\micron\ feature can likely be described with only two parameters.
\citet{Shao18} see the same correlation between the central wavelength and width even though their fit used a symmetric Gaussian.
The variation in the 10~\micron\ central wavelength and width could be caused by variations in the composition and/or size of the silicate grains \citep{Day79, Dorschner86, Dorschner95}.
It was not possible to test if this is the case also for the 20~\micron\ feature, because the width of that feature was fixed in the fitting.
The lower right plot shows that the 10 and 20~\micron\ feature strengths are somewhat correlated, but with scatter.
The significance of this correlation is difficult to judge as we were not able to measure all the 20~\micron\ feature parameters for all our sightlines due to limited wavelength coverage.

Overall, the dust grain models have 10~\micron\ silicate feature properties that are similar to the diffuse average except that their central wavelengths are all too short.
All the models have weaker 20~\micron\ silicate features than the diffuse average.

\subsection{Dense Sightlines}

\begin{figure*}[tbp]
\epsscale{1.15}
\plotone{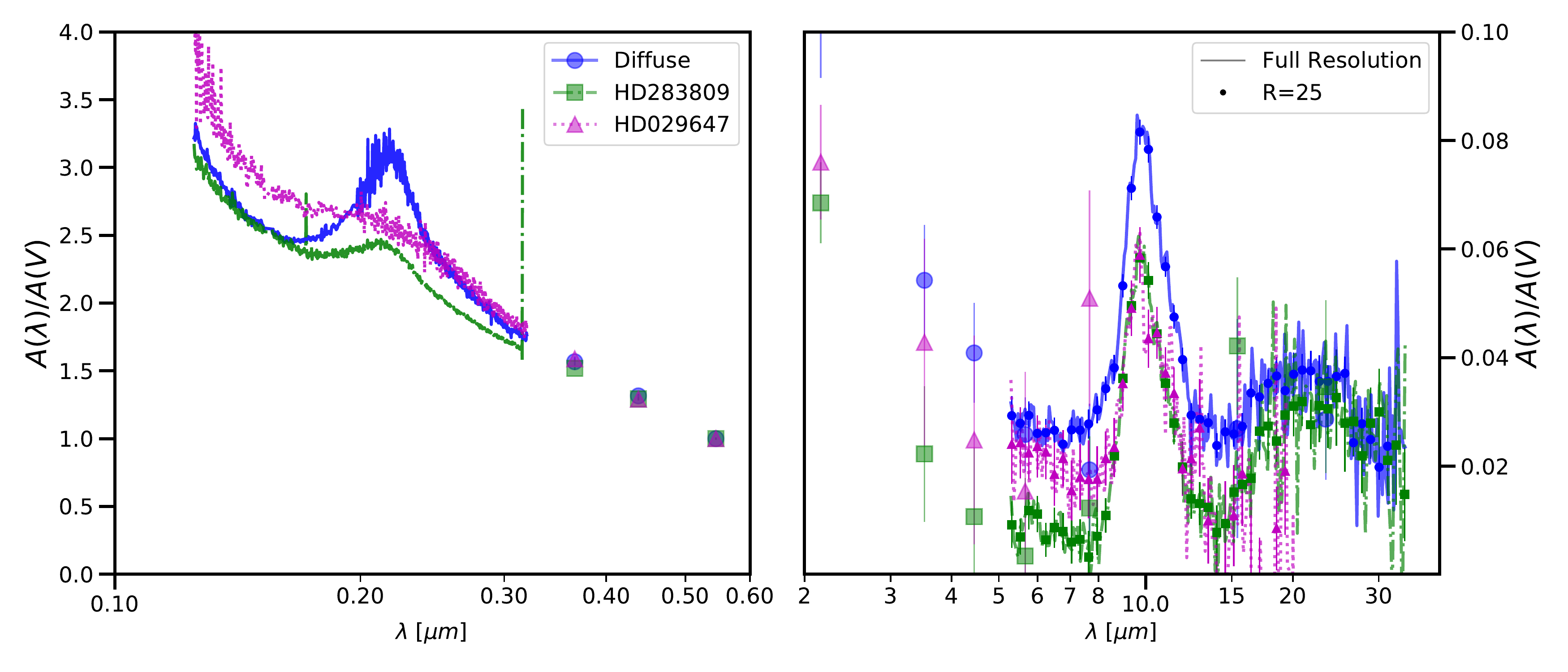}
\caption{The two sightlines that show ice absorption indicative of colder and denser sightlines are shown along with the diffuse average.
The MIR and UV extinctions for both dense sightlines are markedly different than the diffuse average generally showing weaker continuum and silicate and 2175~\AA\ feature extinctions.
The IRS portion of the curve is shown at full resolution and rebinned at a resolution of $R=25$.
\label{fig_diffuse_aveext_wdense}}
\end{figure*}

There are only two sightlines in our sample that clearly probe colder and denser sightlines due to the detection of 3~\micron\ ice absorption.
It is interesting to compare the MIR and UV extinction curves for these two sightlines (Fig.~\ref{fig_diffuse_aveext_wdense}) even though a sample of two is not statistically significant.
The MIR extinction for both dense sightlines is weaker than the diffuse average.
The 10~\micron\ silicate feature strength is smaller than the diffuse average, but still within the distribution of the diffuse sample (Fig.~\ref{fig_silcor}) as are the other properties of this feature.
Intriguingly, the 2175~\AA\ bump strength is much weaker than the diffuse average and well outside of the distribution for the diffuse sample (Fig.~\ref{fig_silcor}).
A much larger sample of dense sightlines with measured MIR and UV extinction curves is needed to determine if these differences are common to dense sightlines.

\subsection{Faint Features Search}
\label{sec_faint_features}

\begin{figure*}[tbp]
\epsscale{1.1}
\plotone{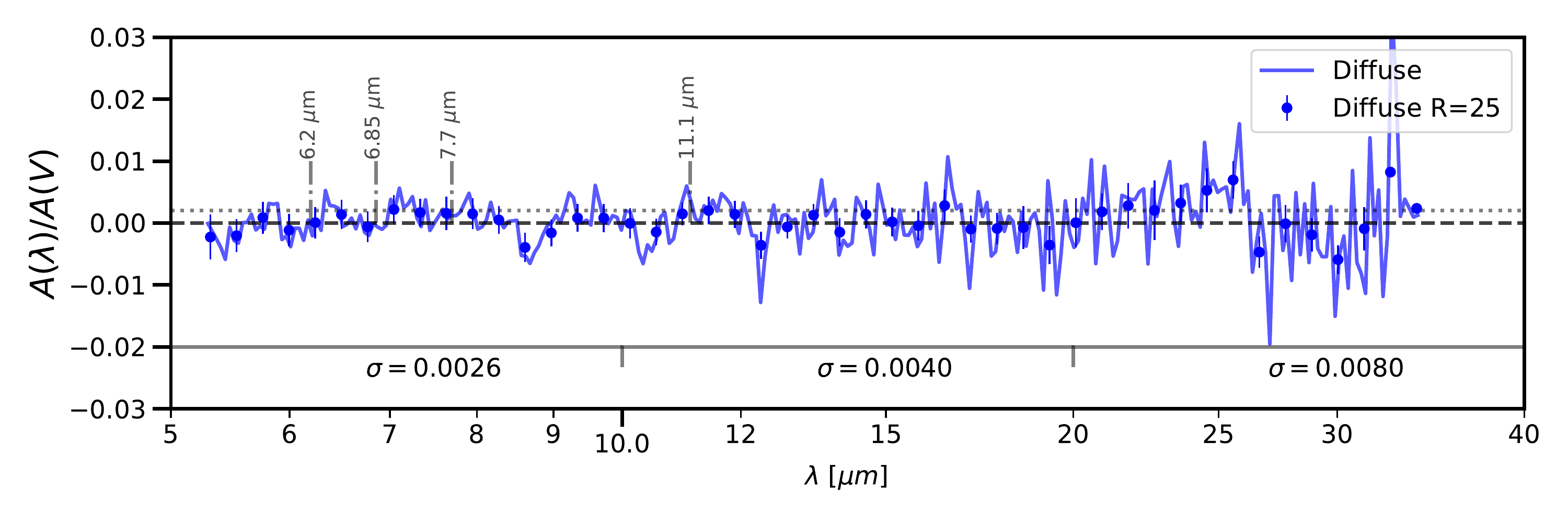}
\caption{The residuals remaining after subtracting the G21 fit from the average diffuse extinction curve are shown.
The wavelengths of the three features detected by \citet{Hensley20} at 6.2, 6.85, and 7.7~\micron\ along the Cyg~OB2~12 sightline are indicated with the dotted horizontal line giving their approximate peak strength of 0.002~\alav.
The wavelength of the crystalline silicate feature detected by \citet{Do-Duy20} at 11.1~\micron\ is indicated.
The average residuals for three wavelength regions are shown along the bottom of the plot.
\label{fig_resid}}
\end{figure*}

We searched for features other than the 10 and 20~\micron\ silicate features by examining the residuals of the G21 fit (Sec.~\ref{sec_normalization}) to the diffuse average curve (Fig.~\ref{fig_resid}).
These residuals provide the best case for finding faint features as the average has an effective $A(V) \sim 42$ and averages over sightline variations.
The lack of large scale structure in the residuals shows that the G21 functional form comprising of a power law with two modified Drude profiles is a good analytic description of the MIR extinction curve.
The lack of sub-structure in the 10~\micron\ feature is expected as the diffuse ISM silicate grains have been shown to be dominated by amorphous material \citep{Kemper04}.
We put statistical limits on the strength of faint features in our diffuse average in three wavelength regions.
Using the residuals, we calculate the sigma upper limits on feature strengths in \alav\ units of 0.0026 (5--10~\micron), 0.0040 (10--20~\micron), and 0.0080 (20--40~\micron).

\citet{Hensley20} found evidence for three features at 6.2, 6.85, and 7.7~\micron\ in the sightline towards Cyg~OB2~12 that has $A(V) \sim 10$.
For this same sightline with the same data, \citet{Potapov20} confirm the detections at 6.2 and 6.85~\micron\ but not the 7.7~\micron\ detection.
The locations of these features and approximate strengths are shown in Fig.~\ref{fig_resid}.
We do not detect these features in our diffuse average.
This could be due to our sightlines measuring a lower dust column of $A(V) \sim 3$ on average,
systematic uncertainties in our lower signal-to-noise observations, or there is something special about the Cyg~OB2~12 sightline (e.g., $A(V) \gg 3$ without detected 3~\micron\ ice absorption).

We also searched for structure near 11.1--11.2~\micron\ in the wing of the silicate feature that might signal the presence of a crystalline silicate component \citep{Do-Duy20} or silicon carbide \citep{Whittet90}.
In the diffuse ISM, \citet{Do-Duy20} find the 11.1~\micron\ feature has an amplitude that is $\sim$5\% of the 10~\micron\ silicate feature.
This gives an amplitude of $A(11.1~\micron)/A(V) \sim 0.003$ based on the diffuse ISM 10~\micron\ silicate amplitude given in Table~\ref{tab_mir_sil_ext_params}.
We found a possible weak feature at 11.1~\micron\ at approximately this level, but caution that this is a tentative detection at best.
The 11.1~\micron\ feature strength is near our detection threshold and similar sized residuals are seen at other wavelengths (e.g., $\sim$9.5~\micron).

\subsection{$R(V)$ and $A(V)$ Prescriptions}

\begin{figure*}[tbp]
\epsscale{1.1}
\plotone{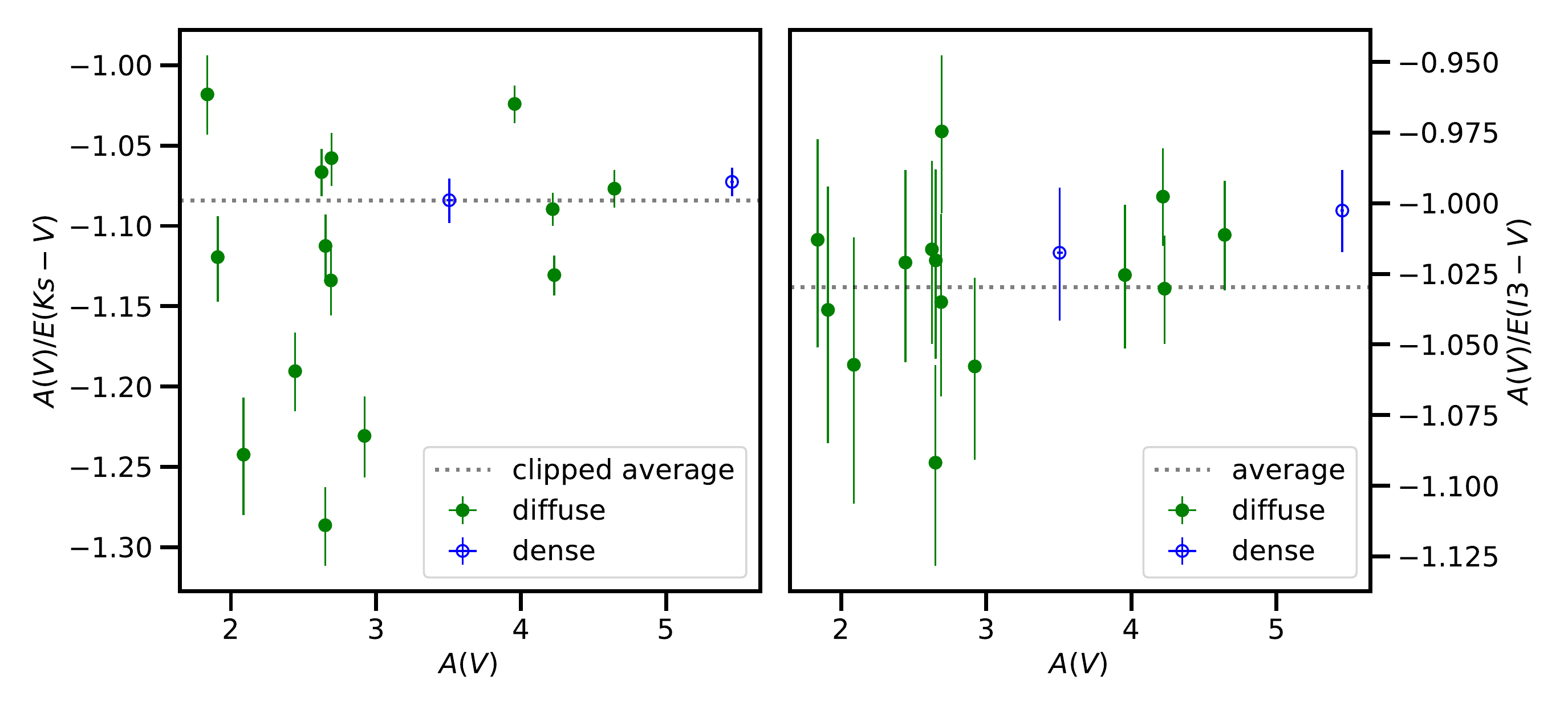}
\caption{The ratios of $A(V)$ to $E(K_s-V)$ and $E(I3-V)$ are shown versus $A(V)$.
The value of $A(V)/E(K_s-V)$ shows strong deviations from the clipped average value of -1.08 where the clipping removed the four points with $A(V)/E(K_s-V) < -1.17$.
The $A(V)/E(I3-V)$ shows a tight correlation around the average value of -1.032.
Note that the $A(V)$ uncertainties are plotted and they are all on the order of the symbol size or smaller.
\label{fig_avekv}}
\end{figure*}

One of the usual methods to determine $R(V) = A(V)/E(B-V)$ is to calculate it by multiplying $E(K-V)/E(B-V)$ by a constant.
This is based on $E(\lambda - V)/E(B-V)$ becomes $-A(V)/E(B-V)$ at infinite wavelength where $A(\lambda=\infty) = 0$ and $A(K)$ is ``near'' infinite wavelength.
The asymptotic behavior of extinction curves at longer wavelengths can be visually seen in Fig.~\ref{fig_ir_elv}.
Calculating $R(V)$ this way is equivalent to calculating $A(V)$ by multiplying $E(K - V)$ by the same constant since
\begin{equation}
R(V) = C \frac{E(K - V)}{E(B - V)} = \frac{A(V)}{E(B - V)}
\end{equation}
and thus
\begin{equation}
A(V) = C E(K - V) .
\end{equation}
Values of $C$ in the literature range from -1.10 to -1.13 \citep{Rieke85, Fitzpatrick99}.
We have derived $A(V)$ in this work using the technique of fitting observations from 1 to 20 -- 38~\micron.
This should result in $A(V)$ values that are of high fidelity as the measurements used are much closer to the value at infinite wavelength than the $K_s$ band.
We tested the simple method of using $E(K_s-V)$ to calculate $A(V)$ in Fig.~\ref{fig_avekv} and found strong deviations from the clipped average value of -1.08 with the clipped sample having a standard deviation of 0.04 or $\sim$4\%.
The non-clipped weighted average is -1.14 and has a standard deviation of 0.08 or $\sim$7\%.
Longer wavelength photometry may provide a better correlation with $A(V)$ and we found that the IRAC3 photometry provided the best correlation with $A(V)$ as is illustrated in Fig.~\ref{fig_avekv} with a value of C = -1.03 and a standard deviation of 0.03 or $\sim$3\%.
It is likely $E(I3 - V)$ is best correlated with $A(V)$ as it is the longest wavelength photometry that is not affected by the silicate features (see Fig.~\ref{fig_diffuse_aveext}).

\section{Conclusions}
\label{sec_conclusions}

We present the mid-infrared Spitzer-based photometric {\em and} spectroscopic extinction curves for a sample of Milky Way sightlines.
The sample is representative of Milky Way extinction sightlines with $A(V)$ = 1.8--5.5~mag, $R(V)$ = 2.4--4.3, and most have measured ultraviolet extinctions.
While most of the sample probes diffuse sightlines, two dense sightlines with 3~\micron\ ice detections are included.
The 10~\micron\ silicate feature was detected along all sightlines even at $A(V) = 1.8$, close to the lowest dust column \citep[$A(V) \sim 1$,][]{Poteet15} for which this feature has been detected in Milky Way extinction curves.

We find that:

\begin{itemize}
\item The NIR and MIR extinction curve can be well fit with a combination of a power law and two modified Drude profiles.  We introduced the modified Drude profile to allow for variably asymmetric profiles.
\item The 10~\micron\ and 2175~\AA\ extinction feature strengths are not correlated -- for the first time providing direct empirical evidence that these two features arise from different grain populations.
\item The average diffuse extinction for our sample is weaker than most, but not all published MIR extinction measurements.  This adds to the evidence of real MIR extinction variations in the Milky Way.
\item The average diffuse extinction is stronger than predicted by existing dust grain models.
\item The residuals between the average curve and its analytic fit were searched for fainter features with wavelength dependent upper limits reported.
\item Finally, we compared our values of $A(V)$ derived from the analytic fits to the full NIR--MIR extinction curve with those computed from prescriptions based on single colors.
Using the $IRAC3-V$ color is significantly better than the usual $K_s-V$ color for measuring $A(V)$ and, by definition also $R(V) = A(V)/E(B-V)$.
\end{itemize}

The IRS spectra and extinction curves are available electronically \citep{zenododata}.
The code used for the analysis and plots is available \citep{githubscripts}.

\acknowledgements
We thank Tobias Fritz for a useful discussion about Brackett line strengths.
This work is based on observations made with the {\em Spitzer Space Telescope}, which was operated by the Jet Propulsion Laboratory, California Institute of Technology under NASA contract 1407.
Support for this work was provided by NASA through Contract Number \#960785 issued by JPL/Caltech.
This publication makes use of data products from the Two Micron All Sky Survey, which is a joint project of the University of Massachusetts and the Infrared Processing and Analysis Center/California Institute of Technology, funded by the National Aeronautics and Space Administration and the National Science Foundation.
This research has made use of the SIMBAD database, operated at CDS, Strasbourg, France \citep{Wenger00}.

\software{Astropy \citep{astropy:2013, astropy:2018}; dust\_extinction \citep{dustextinction}; measure\_extinction \citep{measureextinction}}

\appendix

\section{Spitzer Images}
\label{spitzer_images}

\begin{figure*}[tbp]
\epsscale{1.0}
\plotone{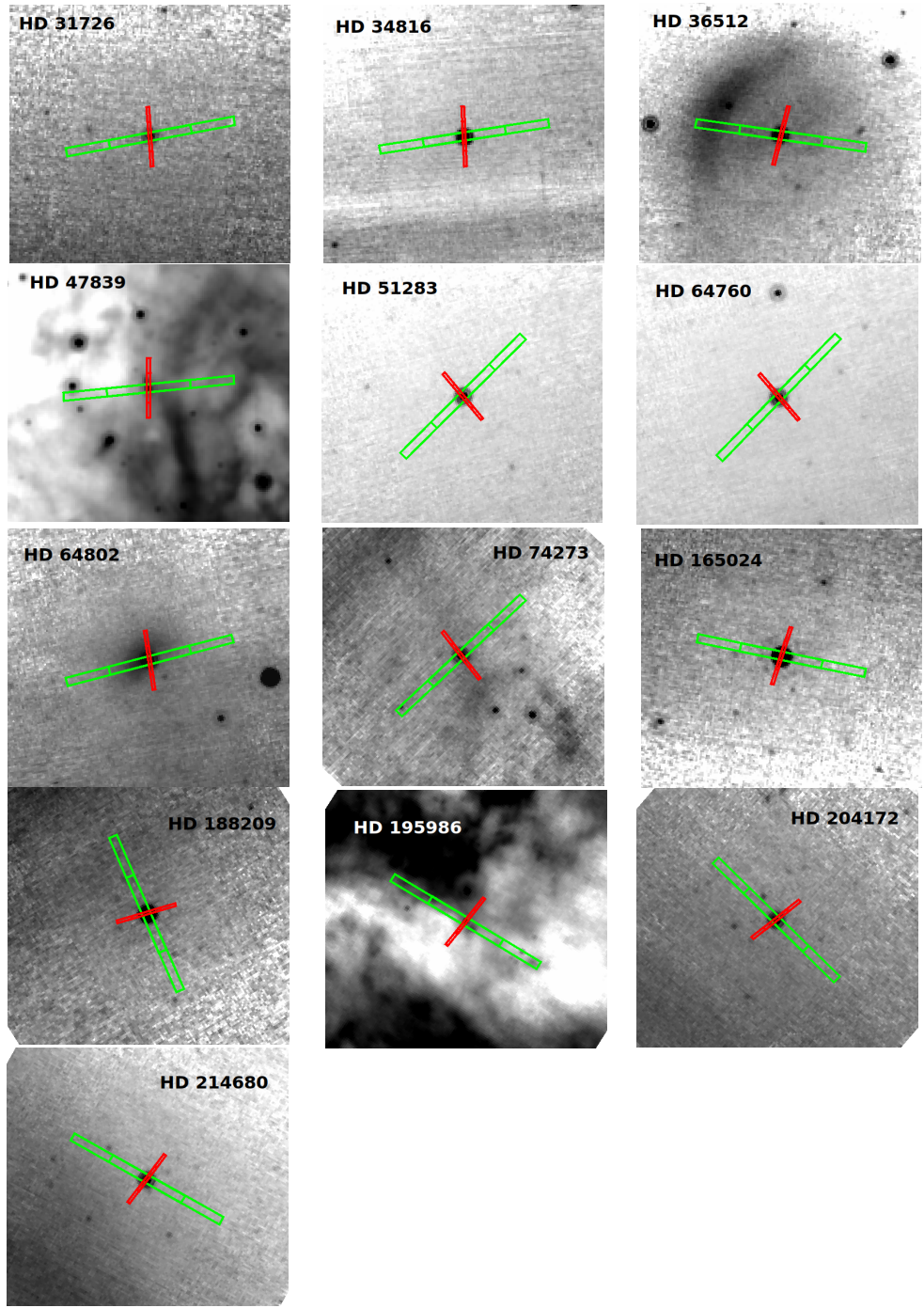}
\caption{The  MIPS 24~\micron\ images of the
  comparison stars are shown.  The location of the IRS SL and LL slits at both nod positions
  are shown in red and green, respectively. Image orientation is north up and east to the left.
\label{fig_images_standards}}
\end{figure*}

The regions around each of the observed stars are shown using final MIPS 24~\micron\ mosaics in Fig.~\ref{fig_images_standards} for the comparison stars and in Fig.~\ref{fig_images_reddened1}--\ref{fig_images_reddened2} for the reddened stars.
These images are displayed stretched to emphasize weak, extended emission and are approximately $5.5\arcmin \times 5.5\arcmin$.
They show that most of our stars are well isolated from nearby sources and extended emission.
This is especially true for the comparison star sample where there is some low level nearby emission for HD47839, HD64802, and HD195986.
For the reddened star sample, the incidence of nearby emission is more prevalent and is seen for HD29647, HD34921, HD96042, HD197702, HD206773, HD229238, HD281159, and Cyg~OB2~8a.
Yet, the presence of nearby emission is not an indication that the source is not a good sightline for extinction curve measurements.
Of the eight reddened stars with nearby emission, only half have clear IRS spectral signatures of strong winds (HD34921, HD206773, \& HD229238) or have poor spectra (HD96042).

It is striking that a few of our sources have wind blown structures surrounding or near them.
The regions nearby BD+63D1964 and HD204827 are particularly nice examples of such structures.

\begin{figure*}[tbp]
\epsscale{1.0}
\plotone{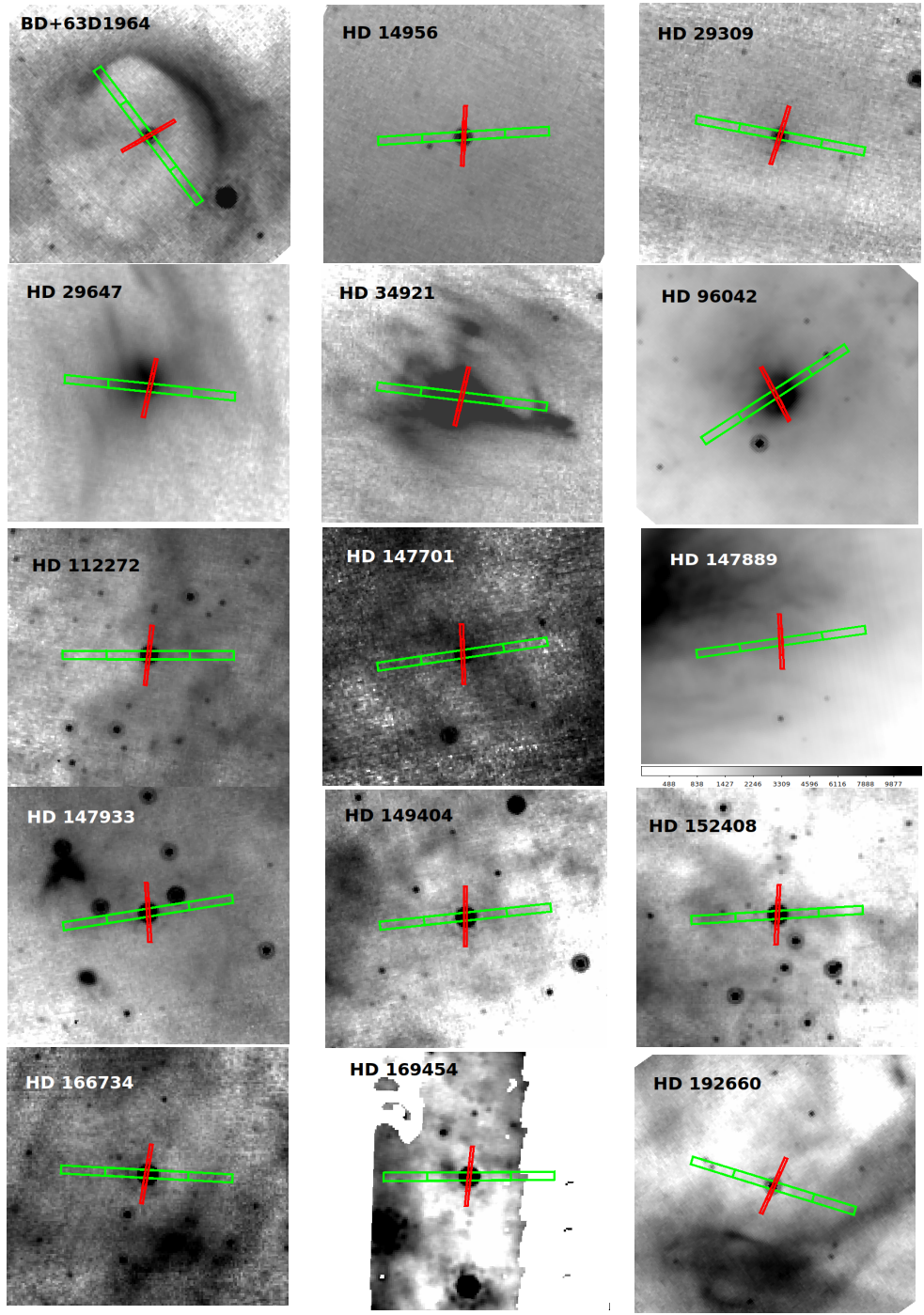}
\caption{As in Fig. \ref{fig_images_standards} for the reddened sample.
\label{fig_images_reddened1}}
\end{figure*}

\begin{figure*}[tbp]
\epsscale{1.0}
\plotone{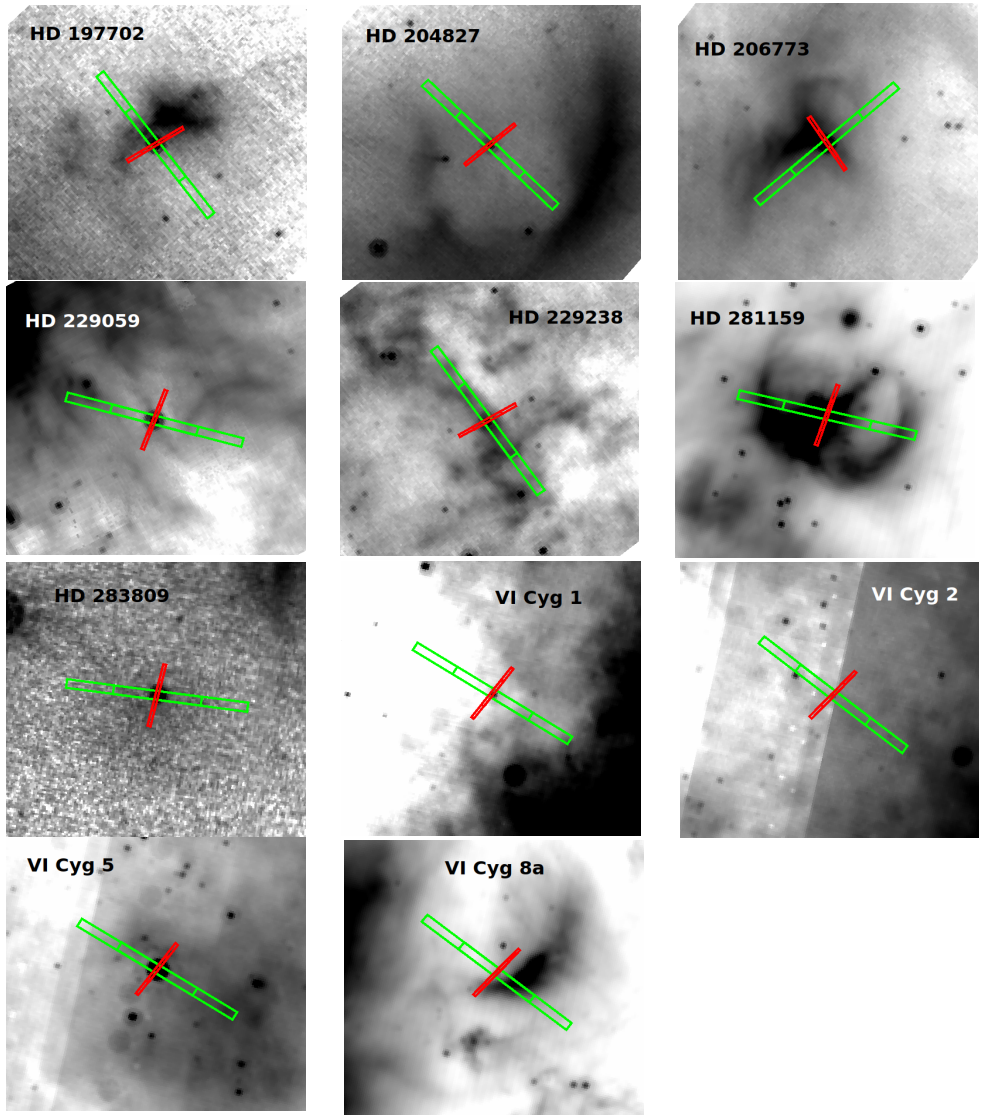}
\caption{Continued from Fig. \ref{fig_images_reddened1} }
\label{fig_images_reddened2}
\end{figure*}

\end{document}